\documentclass[traditabstract]{aa}

\usepackage{graphicx}
\usepackage{txfonts}
\usepackage{natbib}
\bibpunct{(}{)}{;}{a}{}{,} 

\newcommand{\modified}[1]{#1}

\begin{document}

\title{Deep Westerbork observations of Abell 2256 at 350~MHz}
\titlerunning{Abell 2256 at 350~MHz}
\author{M.A. Brentjens}
\authorrunning{Brentjens}
\institute{ASTRON, P.O. Box 2, 7990 AA Dwingeloo, the Netherlands}
\offprints{brentjens@astron.nl}
\date{Received 26 January 2007 / Accepted 26 June 2008}

\abstract
{  Deep polarimetric Westerbork observations of the galaxy cluster
\object{Abell~2256} are presented, covering a frequency range of
\modified{325--377~MHz}. The central halo source has a diameter of the
order of 1.2~Mpc (18\arcmin), which is somewhat larger than at
1.4~GHz. With $\alpha=-1.61\pm0.04$, the halo spectrum between 1.4~GHz
and 22.25~MHz \modified{is less steep} than previously thought. The centre of the
ultra steep spectrum source in the eastern part of the cluster
exhibits a spectral break near 400~MHz.  It is estimated to be \modified{at
least 51 million years old, but possibly older than 125 million years.
A final measurement requires observations in the 10--150~MHz range.} It
remains uncertain whether the source is a radio tail of Fabricant
galaxy 122, situated in the northeastern tip of the source. Faraday
rotation measure synthesis revealed no polarized flux at all in the
cluster.  The polarization fraction of the brightest parts of the
relic area is less than 1\%.  The RM-synthesis nevertheless revealed 9
polarized sources in the field enabling an accurate measurement of the
Galactic Faraday rotation ($-33\pm2$~rad~m$^{-2}$ in front of the
relic). \modified{Based on its depolarization on longer wavelengths, the
line-of-sight magnetic field in relic filament G is estimated to be
between 0.02 and 2~$\mu$G. A value of $0.2$~$\mu$G appears most
reasonable given the currently available data.}
}


\maketitle


\section{Introduction}
\label{brentjens_abell2256_sec:introduction}

Abell~2256 is one of the most massive clusters in the nearby
universe. It has been studied extensively at X-ray, optical, and radio
wavelengths. The galaxies associated with the cluster can be divided
into three distinct populations, based on their kinematics
\citep{BerringtonLuggerCohn2002}. X-ray images of the cluster reveal
several substructures superimposed on a fairly smooth main source
\citep{SunEtAl2002,MolendiEtAl2000, RoettigerBurnsPinkney1995,
BrielEtAl1991}, indicating that Abell~2256 is currently
involved in a major merger event.

The cluster has some interesting radio properties. It contains the
largest number of head-tail galaxies of all known clusters
\citep{MillerOwenHill2003}. One of them has a straight, almost 1~Mpc
long tail.  There is a complex steep spectrum source in the eastern
part of the cluster
\citep{RottgeringEtAl1994,MassonMayer1978,BridleEtAl1979}.  The
cluster is embedded in a large, diffuse, unpolarized radio halo
\citep{BridleFomalont1976, Kim1999, ClarkeEnsslin2006} that dominates
the total flux at decametric wavelengths \citep{CostainEtAl1972}.  The
most striking feature is the large, bright relic source in the
northwestern part of the cluster. Although it is \modified{20--40\%} linearly
polarized at 1.4~GHz \citep{BridleEtAl1979,ClarkeEnsslin2006},
\citet{Jaegers1987} did not detect any linear polarization at
608.5~MHz, establishing a $3\sigma$ upper limit of 20\%
fractional polarization.

There are strong indications that relic sources are caused by large
scale structure formation (LSS) shocks compressing buoyant bubbles of
magnetized plasma that have been emitted by active galactic nuclei in
the past \modified{\citep{EnsslinGopalKrishna2001, EnsslinBrueggen2002,
  Brueggen2003, HoeftBrueggen2007}}.  Shortly after their injection
into the intergalactic medium (IGM), these bubbles have faded beyond
detection limits due to synchrotron losses and adiabatic expansion. An
encounter with a LSS shock wave would then compress the plasma,
increasing and aligning the magnetic field, and accelerate the
electrons to energies enabling synchrotron emission at radio
wavelengths. This model explains both the high fractional polarization
that is usually observed in these relic sources and their peripheral
position.

De Bruyn \& Brentjens (2005) \nocite{DeBruynBrentjens2005} have
discovered several highly polarized structures in the direction of the
\object{Perseus cluster} that were tentatively attributed to the
cluster itself. Some of the structures resembled buoyant bubbles in
the IGM of the Perseus cluster, while others looked like long,
straight shock fronts. One of these "fronts" was located at the
western edge of the cluster at the interface to the Perseus-Pisces
supercluster, exactly where such shocks are expected to occur.
However, these structures have not been detected in Stokes $I$, hence
the exact fractional polarization is unknown. The two most likely
reasons for the non-detection of Stokes $I$ are:
\begin{itemize}
\item The sources are highly polarized and the Stokes $I$ brightness
is below the confusion limit of the Westerbork Synthesis Radio
Telescope (WSRT) in these observations;
\item The structures have Stokes $I$ structure at much larger scales
than the Stokes $Q$ and $U$ structure, rendering it invisible at even the
shortest interferometer baseline.
\end{itemize}  
The latter situation is common in the Galactic synchrotron foreground
\citep{Wieringa1993}. A Galactic interpretation can therefore not be
ruled out. However, \citet{GovoniEtAl2005} have observed similar
structures in Stokes $I$ near \object{Abell~2255} with the VLA at
1.4~GHz. Abell~2256 was chosen to search for similar objects
because of its proximity, its dynamic nature, its relic source, and
its moderate Galactic latitude of $32\degr$.

Wide field polarimetric images with noise levels well below
1~mJy~beam$^{-1}$ are required in order to detect such sources.
Additionally, the images can be used to find more polarized background
sources in order to establish the Galactic contribution to the Faraday
rotation measure (RM) towards the cluster, and \modified{to} determine the
decline of the fractional polarization with increasing wavelength,
enabling a measurement of the thickness of the relic sources in
RM-space.

\modified{The spectrum} of the diffuse halo source has been rather uncertain.
\citet{BridleEtAl1979} favoured a spectral index
($S_\nu\propto\nu^\alpha$) \modified{of $\alpha < -1.2$ between 610 and 1415~MHz
and $\alpha \approx -1.8$ between 151 and 610~MHz}, based on maps where
the halo was only marginally detected. More recently
\citet{ClarkeEnsslin2006} determined the 1.4~GHz halo flux in more
sensitive VLA D \modified{observations, but could not derive the spectrum of the
halo because of the large uncertainties in the halo flux measurements
available to them}. They \modified{assumed} that $-1.7 \lesssim \alpha \lesssim
-1.25$. \citet{Kim1999} determined that $\alpha=-2.04\pm0.04$, but
this estimate was partly based on an erroneous value of the total
cluster flux at 81.5~MHz, and partly on maps in which the halo was
only marginally detected. New measurements of the 351~MHz flux of the
halo and the entire cluster are presented in
Sect.~\ref{brentjens_abell2256_sec:maps}.  These values are combined
with corrected literature values in order to obtain accurate spectra
for the cluster as a whole and the diffuse halo source itself.

The redshift of \object{Abell~2256} is 0.058
\citep{BerringtonLuggerCohn2002}. Assuming $H_0 =
71$~km~s$^{-1}$~Mpc$^{-1}$ \citep{SpergelEtAl2003}, 1\arcmin\ on the
sky corresponds to 67.8~kpc at the cluster.



\section{Observations}
\label{brentjens_abell2256_sec:observations}

The observations were conducted with the Westerbork Synthesis Radio
Telescope, which consists of fourteen parallactic 25~m
dishes on an east-west baseline and uses earth rotation to fully
synthesize the uv-plane in 12~h. There are ten fixed dishes
(RT0--RT9) and four movable telescopes (RTA--RTD). The distance
between two adjacent fixed telescopes is 144~m.

The data set consists of two observing sessions of 12~h each. The first
during the night of May 17/18 2004 with RT9--RTA = 36~m and the
second during the night of May 18/19 with RT9--RTA = 72~m.  The
distances between the movable dishes were kept constant (RTA--RTB =
RTC--RTD = 72~m, RTB--RTC = 1224~m).  The uv-plane is therefore
sampled at regular intervals of 36~m out to the longest baseline of
2736~m. Only the zero spacing is missing. The regular interval causes
a grating ring with a radius of 80$\arcmin$\ at 350~MHz. At this
frequency the $-5$~dB and $-10$~dB points of the primary beam are at
radii of $70\arcmin$\ and $120\arcmin$\ respectively. The observations
are sensitive to angular scales up to $90\arcmin$\ at a resolution of
$67\arcsec$\ full width at half maximum (FWHM).

The eight frequency bands were each 10~MHz wide and centred at 319,
328, 337, 346, 355, 365, 374, and 383~MHz. The WSRT is equipped with
linearly polarized feeds for this frequency range.  The $x$ dipole is
oriented north-south and the $y$ dipole  east-west.  The correlator
produced 64 channels in all four cross correlations for each band with
an integration time of 30~s. The on-line system applied a uniform
lag-to-frequency taper. A Hanning lag-to-frequency taper was applied
off-line, effectively halving the frequency resolution.  The pointing
centre and phase centre were at the optical centre of the cluster at
$\alpha = 17^\mathrm{h}03^\mathrm{m}45^\mathrm{s}$, $\delta =
+78\degr43\arcmin00\arcsec$\ (J2000.0). 

The observations were bracketed by two pairs of calibrators, each
consisting of one polarized and one unpolarized source.
\object{3C~295} and the eastern hot spot of \object{DA~240} were
observed before Abell~2256 and \object{PSR~B1937+21} and
\object{3C~48} afterwards.

The theoretical image noise in one Hanning tapered channel of a single
12~h observation at 350~MHz is about 1.56~mJy~beam$^{-1}$ (uniform
weighting).  Band 1 was unusable on the second night due to a
polarization calibration problem. Band 4 suffered from a strange,
broad band interference on the shorter baselines and was therefore
discarded.  Band 8 (383~MHz) was unusable in both nights due to man
made interference. The four lowest and four highest channels of the
remaining bands were discarded. Only the odd channels were selected
because they are linearly dependent on the even channels due to the
Hanning taper. In the remaining 28 channels per band, 60\%  of the
data was usable.  The expected thermal noise level in the integrated
polarization maps is therefore
1.56~mJy~beam$^{-1}$/$\sqrt{2\times28\times5\times0.6} \approx
0.12$~mJy~beam$^{-1}$. This level can not be reached in the total
intensity maps because they are confusion limited at approximately
0.3~mJy~beam$^{-1}$ at this frequency and resolution
\modified{\citep{WSRTGuide}}.


\section{Data reduction}
\label{brentjens_abell2256_sec:datareduction}

Flagging, imaging, and self calibration were performed with the AIPS++
package \citep{McMullinGolapMyers2004}. System temperature
corrections, flux scale calibration, polarization calibration,
ionospheric Faraday rotation corrections, and deconvolution were
performed with a calibration package written by the author and based
on the AIPS++ and CASA libraries.

\subsection{System temperatures}

The system temperature readings of the 36~m observing session were
unfortunately unusable due to problems with the preparation of the
telescope for the observations.  The system temperature readings of
the 72~m session were usable, but had large spikes caused by RFI.  The
system temperature corrections were therefore only applied to the 72~m
data, not to the 36~m data. The 72~m system temperatures were filtered
using a median window with a total width of 20~minutes, which was very
effective in mitigating the effect of short bursts of RFI.  The
difference between the $x$ and $y$ system temperatures of some bands
in several antennas was of the order of 30--50~K, although for most
band/antenna combinations it was less than 5~K. Because of the lack of
system temperature data in the 36~m session, the 72~m data were used
to determine the flux scale of the point source model used for self
calibration. The 36~m data were later tied to the same flux scale in
order to combine them with the 72~m data for the Stokes $I$ image.
The 36~m data were not used for polarimetry because there was small,
but significant polarization leakage even after cross-calibration.
This may have been caused by temporal variations in the difference in
the system temperatures between the $x$ and $y$ receptors during the
observation.

\subsection{Bandpass- and polarization  calibration}
\label{brentjens_abell2256_sec:unpolarized}

The flux scale and polarization leakages were calibrated
simultaneously using the unpolarized calibrator sources
3C~295 and 3C~48. The Measurement Equation
\citep{HBS2} for the visibility on the baseline $i$--$j$ at a certain
frequency is given by 
\begin{equation} 
\tens{\tilde{V}}_{ij} = \tens{J}_i \tens{V}_{ij} \tens{J}_j^\dagger,
\end{equation}
where
\begin{equation}
\tens{V}_{ij} = \left(\begin{array}{cc}
<xx> & <xy> \\
<yx> & <yy>
\end{array}\right)_{ij}
\end{equation}
represents the pure sky visibilities in terms of the cross
correlations between the $x$ and $y$ receptors of the antennae in the
baseline, $\tens{J}_i$ and $\tens{J}_j$ are the Jones matrices
describing the properties of the antennae, and $\tens{\tilde{V}}_{ij}$
is the observed coherency matrix. The $\dagger$ denotes Hermitian matrix
transposition.

The fluxes of the calibrator sources were computed using the
expressions and coefficients given in \citet{PerleyTaylor1999}\footnote{Note
that the 327~MHz flux scale of WSRT observations has since 1985 been
based on a 325~MHz flux of 26.93~Jy for \object{3C~286} (the
\citet{BaarsEtAl1977} value). On that flux scale, the 325~MHz flux of
\object{3C~295} is 64.5~Jy, which is almost 7\%\ higher than the value
assumed at the VLA and in this paper (A.~G.~de~Bruyn, private
communication).}, which
extends the \citet{BaarsEtAl1977} flux scale to lower frequencies.
Because each channel is individually tied to this flux scale, all sources
appear in the images with their true spectral indices.

The on-axis and off-axis polarization leakages at the WSRT  are
strongly frequency dependent with a period of the order of 17~MHz.
The leakage amplitude can increase from almost negligible to 1.5\% and
decrease back to negligible over a frequency interval of approximately
8.5~MHz.  It was therefore necessary to solve for all elements of the
Jones matrices for each channel individually. The phases of the
diagonal elements of the Jones matrix of RT1 were fixed at 0.  The
remaining $x$--$y$ phase difference, 
\begin{equation}
\delta_{xy} = \tan^{-1}\frac{V}{U} + n\pi,
\label{brentjens_abell2256_eqn:xy_phase_diff}
\end{equation}
where $n$ is an integer, rotates Stokes $U$ into Stokes $V$. It is
fortunately fairly constant across a 10~MHz band.
Equation~(\ref{brentjens_abell2256_eqn:xy_phase_diff}) has only two
unique solutions: one when $n$ is even, the other when $n$ is odd.
One solution rotates the $(U,V)$ vector to positive $U$, the other to
negative $U$. In order to select the correct solution, one only
requires the sign of the RM of the calibrator sources. The
(sinusoidal) Stokes $U$ spectrum is shifted by 90\degr\ towards
smaller $\lambda^2$ with respect to the Stokes $Q$ spectrum if the RM
is positive and 90\degr\ in the opposite direction if it is negative.

PSR~1937+21 and the eastern hot spot of DA~240 both
have a positive RM. According to \citet{Tsien1982} the RM of
DA~240 is +2.4~rad~m$^{-2}$ (no error quoted). According to
A.G.~de~Bruyn, the RM of DA~240 plus a minimal ionosphere
should be $+3.7\pm0.5$~rad~m$^{-2}$. Using the low frequency front
ends of the WSRT, he found that the RM of the pulsar, also without
correcting for the ionospheric RM, is $+8.5\pm0.5$~rad~m$^{-2}$
(private communication).

\begin{table}
\caption{\modified{Faraday rotation measures and position angles (north through
east) at $\lambda^2 = 0$ for the polarized calibrator
sources.}}
\label{brentjens_abell2256_tbl:calibrator_rm}
\begin{center}
\begin{tabular}{lll}
\hline
\hline
Source & Rotation measure & Position angle\\
       & (rad~m$^{-2}$ )   & (\degr)\\
\hline
\object{PSR~1937+21}& $+7.86\pm0.20$& $291.4\pm1.1$\\
\object{DA~240}     & $+3.33\pm0.14$& $122\ \ \ \pm3$\\
\hline
\end{tabular}
\end{center}
\end{table}

The most accurate rotation measures for these objects to date are
listed in Table~\ref{brentjens_abell2256_tbl:calibrator_rm}. They were
obtained after correcting for ionospheric Faraday rotation using the
procedure outlined in
Sect.~\ref{brentjens_abell2256_sec:ionospheric_rm}.  Because the
shorter spacings were all affected by solar interference, the
polarization angle at each frequency was computed based on the average
visibility of all spacings larger than 700~$\lambda$.

\subsection{Phase calibration}
\label{brentjens_abell2256_sec:selfcal}

After the data were corrected for the more or less time independent
effects described in the previous section, three phase-only self
calibration iterations were performed  on the 72~m data set
\citep{PearsonReadhead1984}. The sky model consisted of a list of 103
bright point source components with accurate positions and $I$, $Q$,
$U$, and $V$ fluxes for each channel, supplemented with a grid of
15525 CLEAN components \citep{Hogbom1974} in Stokes $I$. The CLEAN
model was given a spectrum proportional to $\nu^{-1}$. Both models
were updated after every self calibration step.

The initial models were obtained by:

\begin{enumerate}
\item Making an image of a channel close to 350~MHz with all baselines
$> 600\ \lambda$;
\item Identifying all point sources brighter than 30~mJy~beam$^{-1}$;
\item Solving for positions of all 103 sources in band 5 in the
uv-plane (band 5 had very little RFI);
\item Solving for $I$, $Q$, $U$, and $V$ of all 103 point sources for
each channel in the uv-plane;
\item Solving for phases using the point source model and all baselines
$> 600\ \lambda$ ;
\item Correcting the visibilities for these phases;
\item Subtracting the point source model from the visibilities;
\item Making (residual) Stokes $I$ images of all usable channels and CLEAN them;
\item Constructing the CLEAN model by averaging the CLEAN models of all
channels.
\end{enumerate}

One self calibration iteration consisted of the following
steps:

\begin{enumerate}
\item Adding the point source model to the visibilities
\item Solving for phase corrections using the point source model and the
CLEAN model on all baselines $> 200\ \lambda$;
\item Solving for point source model fluxes;
\item Subtracting the point source model from the visibilities;
\item Making (residual) Stokes $I$ images of all usable channels and CLEAN them;
\item Constructing the CLEAN model by averaging the CLEAN models of all
channels.
\end{enumerate}

The final CLEAN model was the average of the CLEAN models of
individual channel maps that were produced after subtracting the point
source component model.  This lead to a very detailed and smooth
representation of the extended emission in the CLEAN model.

After the self calibration, data points were flagged if the absolute
value of the residual visibility (corrected data minus model data) was
larger than 5 times the root mean square (RMS) amplitude of the residuals
for that band and baseline.

\subsection{Ionospheric Faraday rotation}
\label{brentjens_abell2256_sec:ionospheric_rm}

The ionospheric Faraday rotation at the WSRT is typically 
between 0.2 and a few rad~m$^{-2}$. Daytime values of 5~rad~m$^{-2}$
are possible during solar maximum. A difference of 2 rad~m$^{-2}$ in
ionospheric Faraday depth corresponds to a change in polarization
angle of $> 100\degr$ at the lowest frequency ($\approx 315$~MHz).
The polarized calibrator sources showed a difference of up to 20\degr\
in their polarization angle between the 36~m and 72~m sessions at the
lowest frequencies, even though they were observed at the same time of
the day on consecutive days. It was therefore necessary to compensate
for the ionospheric RM in order to avoid depolarization.

\citet{EricksonEtAl2001} showed that a simple
ionospheric model fed with data from dual band GPS receivers installed
at the VLA could predict the ionospheric rotation measure in most
circumstances to better than 0.04~rad~m$^{-2}$. Sometimes, however,
they observed an unexplained discrepancy of the order of 30\degr\
between the predicted and observed polarization angle of
\object{PSR~1932+109}. This corresponds to a mismatch in RM of
approximately 0.6~rad~m$^{-2}$. 

Because there are only very few lines of sight through the ionosphere
with the single dual band GPS receiver installed at the WSRT, the
ionospheric Faraday rotation was computed using global GPS total
ionospheric electron content (TEC) data and an analytical model of the
geomagnetic field. The GPS-TEC data were provided by the Center for
Orbit Determination in Europe (CODE) of the Astronomical Institute of
the university of Bern, Switzerland. The geomagnetic field was
computed using the US/UK World Magnetic Model (WMM)
\citep{MacMillanQuinn2000}, while the ionosphere was modelled as a
spherical shell with a finite thickness and uniform density at an
altitude of 350~km above mean sea level. The geomagnetic field
parallel to the line of sight was evaluated at the points where the
line of sight from the WSRT towards \object{Abell~2256} pierced
through the model ionosphere. The TEC data are published for all even
UTC hours. The expected uncertainty in RM towards Abell~2256,
based on the TEC RMS uncertainty published by CODE, is typically
0.07--0.1~rad~m$^{-2}$ for each data point during the observations.

The ionospheric RM was also tracked using
\object{6C~B165001.3~+791133}, a polarized source at 53$\arcmin$\ from
the phase centre. Its polarized flux is about 15 mJy at 350~MHz,
yielding an apparent polarized flux of approximately 10.1~mJy after
primary beam attenuation. The total Faraday depth as a function of
time is
\begin{equation} 
\phi_\mathrm{tot}(t) = \phi_\mathrm{src} + \phi_\mathrm{ion}(t),
\end{equation}
where $\phi_\mathrm{ion}(t)$ is the time dependent ionospheric Faraday
rotation and $\phi_\mathrm{src}$ is the constant contribution of
6C~B165001.3~+791133.  Based on the thermal noise,
$\phi_\mathrm{tot}$  could be estimated with an accuracy of
approximately 0.26~rad~m$^{-2}$ every two hours. This translates to a
1$\sigma$ error in the polarization plane of approximately $13\degr$\
at 324~MHz.

\begin{figure}
\centering
\resizebox{\hsize}{!}{\includegraphics{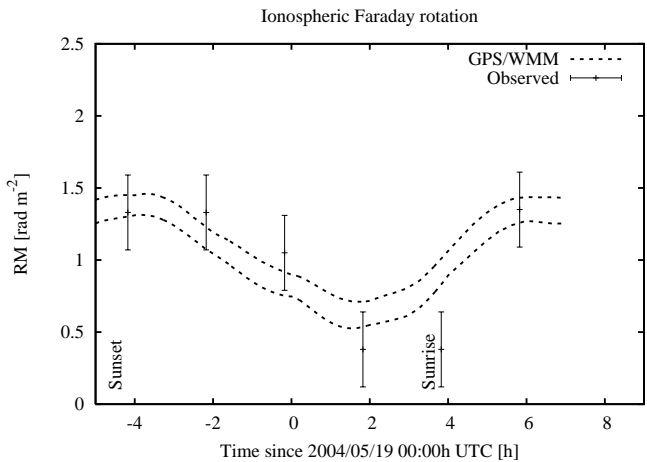}}
\caption{Ionospheric RM as predicted by global GPS-TEC data and the
US/UK World Magnetic Model. The dashed lines indicate the $\pm
1\sigma$ predicted TEC/WMM rotation measures, interpolated between the
2~h intervals at which the data are published. The points with 
error bars are observed RMs of 6C~B165001.3~+791133,
shifted up by 27.0~rad~m$^{-2}$. The horizontal axis is the
time in hours since May 19 2004,~00:00 UTC.}
\label{brentjens_abell2256_fig:ionospheric-rm}
\end{figure}

Figure \ref{brentjens_abell2256_fig:ionospheric-rm} compares the
observed changes in ionospheric Faraday rotation with the predicted
ionospheric RM based on GPS-TEC/WMM data. The \modified{uncertainty in} the
GPS-TEC/WMM \modified{rotation measures is} based on the RMS error in the TEC
along the given line-of-sight.  Uncertainties in the geomagnetic field
are not \modified{incorporated}. After discarding the outlier directly after
sunrise, $\phi_\mathrm{src} = -27.0\pm0.1$~rad~m$^{-2}$.

The \modified{Abell~2256} visibilities were corrected for the
ionospheric Faraday rotation after instrumentally polarized point
sources were \modified{subtracted in order to prevent residuals of these sources
due to distortion of their PSF by the ionospheric Faraday rotation
correction.}

\subsection{Imaging}
\label{brentjens_abell2256_sec:imaging}

The final dirty channel images in all Stokes parameters and the
corresponding point spread functions (PSFs) were created from the
corrected visibilities after subtracting the instrumentally polarized
point sources.  The uv-plane was uniformly weighted. Because of a
fractional bandwidth of 15\%, the area under the main lobe of the PSF
at the lowest frequency is 30\% larger than at the highest frequency,
hence a Gaussian taper was applied to the uv-plane before Fourier
transforming in order to convolve the images to a common resolution of
$67\arcsec$ FWHM.  All maps are in North Celestial Pole (NCP)
projection. The dirty maps are 2048$\times$2048 pixels large, with a
pixel size of $15\farcs6$.  The central 1024$\times$1024 pixels of the
dirty images were deconvolved using a H\"ogbom CLEAN
\citep{Hogbom1974}.

While deconvolving the Stokes $I$ channel images, CLEAN was
constrained to only search for components at certain pixel positions
(the mask). The mask was obtained by averaging all channel maps of a
previous deconvolution, selecting all pixels where Stokes $I$ was
larger than some threshold, and expanding the mask to include all
pixels where at least one of its eight neighbours was part of the
mask. The threshold level was lowered after each \modified{self calibration}
iteration.  The final threshold level for the mask was 1.5
mJy~beam$^{-1}$.  The loop gain was set to 0.3 and the CLEAN was
stopped whenever either 15000 iterations were performed or the maximum
residual in the area where CLEAN components were allowed was less than
0.5~mJy~beam$^{-1}$.  This is roughly one third of the thermal noise
level in the individual channel maps. It was necessary to deconvolve
so deep into the noise of individual channels in order to obtain the
highest possible dynamic range in the average of 116 channel maps. The
CLEAN models were added to the residual images using a circular
Gaussian restoring beam with a FWHM of 67\arcsec. At this declination,
the north-south to east-west ratio of the untapered WSRT PSF is
1.02. The use of a circular restoring beam is justified in this case
because of the strong, circular uv-plane taper that was applied.

After deconvolution, the channel maps were corrected for the total
power primary beam of the WSRT, which can be approximated by
\begin{equation} G(\nu, \gamma) = \cos^6{k \nu \gamma},
\label{brentjens_abell2256_eqn:primary_beam} \end{equation} where
$\nu$ is the frequency in Hz, $\gamma$ is the angular distance from
the pointing centre in radians, and $k = 6.6\times10^{-8}$~s is a
constant, namely the light crossing time across a 19.8~m aperture,
which is roughly the effective diameter of a WSRT dish in this
frequency band (25~m dish, 63\% effective surface area).  Finally, the
best 116 channel maps were averaged.
 
The deconvolution of the Stokes $Q$ and $U$ maps was unconstrained.
The CLEAN was considered complete after 5000 iterations or if the
largest residual on the inner 1024$\times$1024 pixels was less than
1~mJy~beam$^{-1}$. These images were also corrected for the primary
beam. Unfortunately the 36~m observing session suffered from residual
polarization leakage due to gain variations during the observations.
It turned out to be difficult to obtain correct gain solutions, and
system temperature data were corrupted. Therefore the polarization
images were created using only the 72~m spacing.

\subsection{RM-synthesis}
\label{brentjens_abell2256_sec:rmsynthesis}

The low frequency and high fractional bandwidth of these observations
required RM-synthesis \citep{BrentjensDeBruyn2005} in order to avoid
bandwidth depolarization. Images were created for Faraday depths of
-800 to +800~rad~m$^{-2}$ in steps of 2~rad~m$^{-2}$.

Off-axis instrumental polarization was not corrected because it is
limited to very specific Faraday depths after RM-synthesis.  In the
case of the WSRT the off-axis polarization consists of two components:
a largely frequency independent offset and a frequency- and direction
dependent oscillation with a period of 17~MHz.  After RM-synthesis,
the offset ends up at $\phi=0$~rad~m$^{-2}$ and the 17~MHz oscillation
at $|\phi| \approx 44$~rad~m$^{-2}$ (near 350~MHz). The sign of $\phi$
depends on the position angle with respect to the pointing centre.

The noise level in the images ranges from 0.31~mJy~beam$^{-1}$ near
0~rad~m$^{-2}$ to the thermal noise level of 0.12~mJy~beam$^{-1}$ for
Faraday depths $|\phi| > 300$~rad~m$^{-2}$. The polarization images
were corrected for the total intensity primary beam of the WSRT using
Eq.~(\ref{brentjens_abell2256_eqn:primary_beam}) before the
RM-synthesis was applied.


\section{Total intensity}
\label{brentjens_abell2256_sec:maps}

\begin{figure*}
\centering
\includegraphics[width=0.95\textwidth]{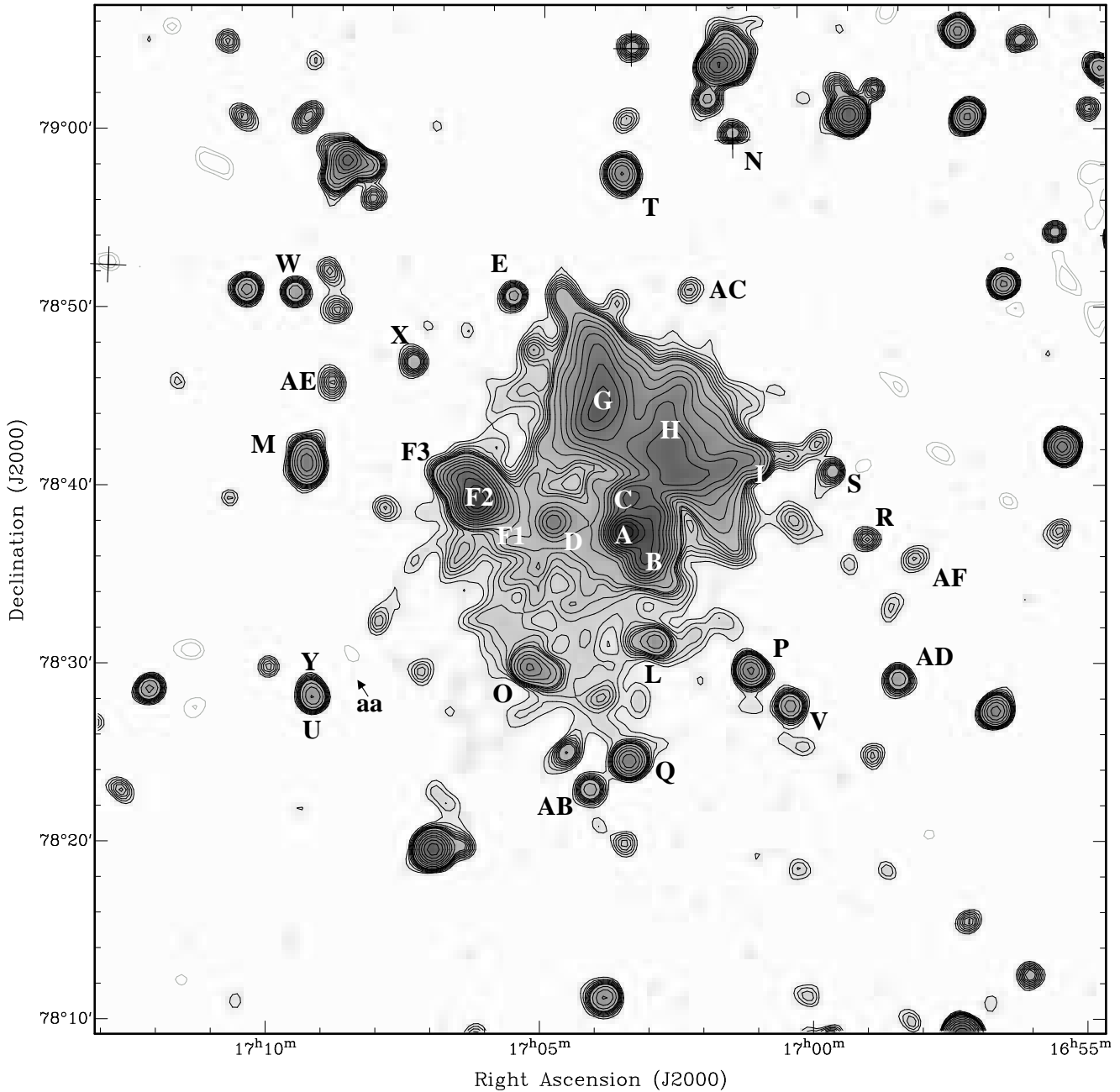}
\caption{\modified{Total intensity image of \object{Abell~2256} observed with
  the WSRT in 2004. Contour levels are at ($-4$, $-3$, 3, 4, 5, 6, 7,
  8, 9, 10, 14, 20, 28, 40, 56, 80, 113, 160, 226,
  320)$\times$0.6~mJy~beam$^{-1}$. The beam is a circular Gaussian
  with a FWHM of $67\arcsec$. Crosses mark locations where point
  source components have been subtracted. The average frequency of
  this multi frequency synthesis map is 351~MHz.} }
\label{brentjens_abell2256_fig:a2256_total_intensity}
\end{figure*}

\begin{figure*}
\centering
\includegraphics[width=0.95\textwidth]{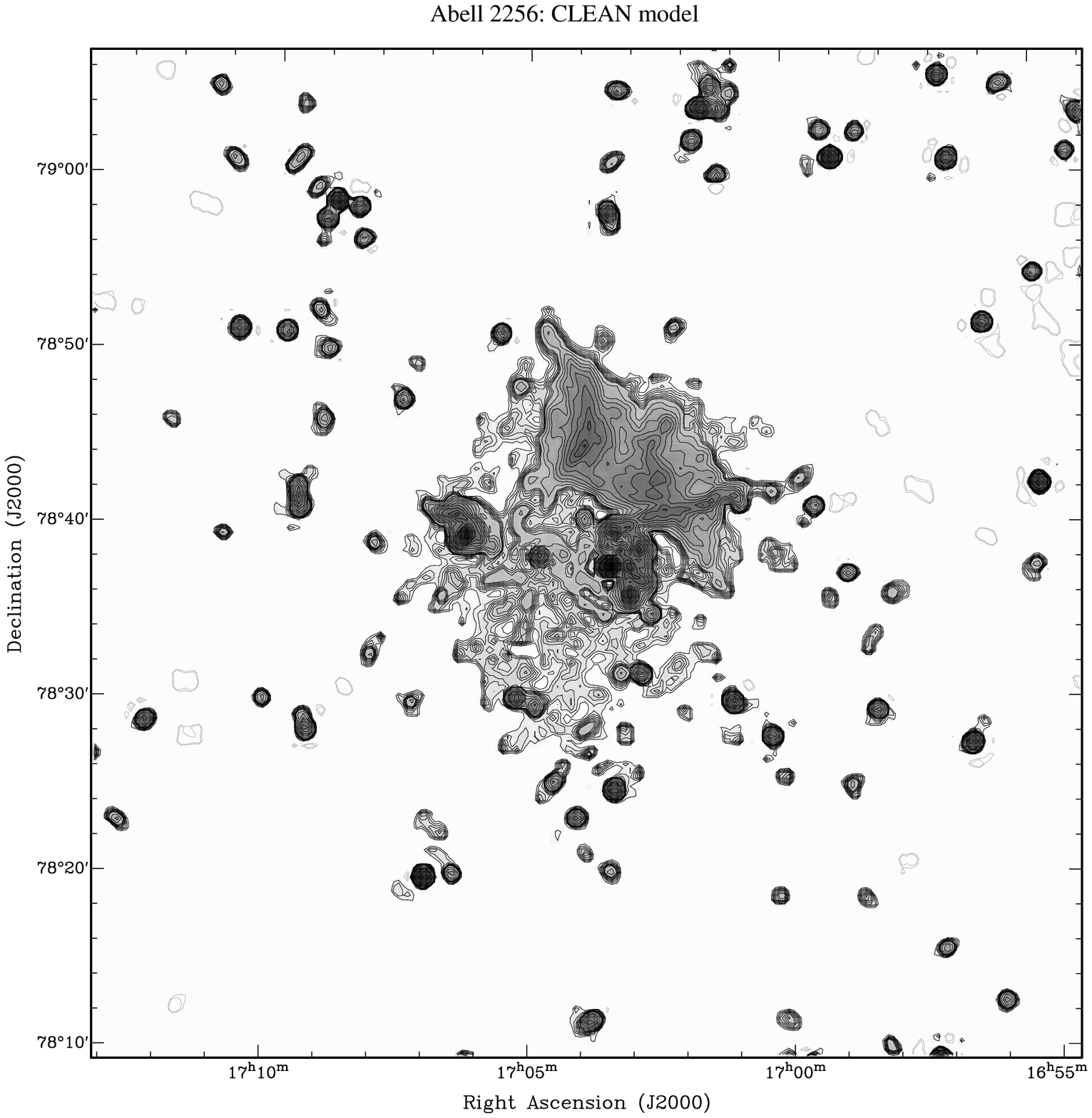}
\caption{The average CLEAN model convolved with a circular Gaussian
  beam with a FWHM of 26\arcsec. The contours are at ($-4$, $-3$, 3,4,
  5, 6, 7, 8, 9, 10, 14, 20, 28, 40, 56, 80, 113, 160, 226,
  320)$\times$91~$\mu$Jy~beam$^{-1}$.}
\label{brentjens_abell2256_fig:cleanmodel}
\end{figure*}

Figure~\ref{brentjens_abell2256_fig:a2256_total_intensity} shows the
central square degree of the final Stokes $I$ image.  The average
frequency of the combined map is 351~MHz and the integrated bandwidth
is 43~MHz spread over  52.5~MHz.  The noise level near the centre of
the map is approximately 0.6~mJy~beam$^{-1}$, dominated by residual
calibration problems. The dynamic range of the full map (brightest
source:central noise level) is of the order of 3000:1.  Sources are
labelled according to \citet{BridleEtAl1979}, including the extension
by \citet{RottgeringEtAl1994}.

Figure~\ref{brentjens_abell2256_fig:cleanmodel} displays a mildly
super-resolved map of the same region. It was obtained by convolving
the CLEAN model with a circular Gaussian beam of $26\arcsec$ FWHM.
This map reproduces the filamentary structure seen in the relic area
by \citep{ClarkeEnsslin2006}. The halo itself appears to exhibit
filamentary structure too.

Except for source AA, all sources mentioned in
\citet{RottgeringEtAl1994} were detected. For source AA, they list a
total flux of 2.8$\pm$0.3~mJy at 1446~MHz. Although they mention the
source has a flux of  8~mJy at 327~MHz, it is not visible in their
327~MHz map.  H.J.A.~R\"ottgering agrees that their flux estimate of
source AA at 327~MHz is caused by an error in the flux estimation
procedure (private communication). My non-detection gives a 3$\sigma$
upper limit of 1.8~mJy at 351~MHz. The source is also invisible in
recent WSRT observations at 150~MHz, yielding a 3$\sigma$ upper limit
of 12~mJy (H.J.A.~R\"ottgering, private communication).  The spectral
index $\alpha$ ($S_\nu \propto \nu^{\alpha}$) between 351~MHz and
1446~MHz must therefore be larger than 0.3.

The remainder of this paper focuses on  the diffuse halo emission
that pervades the cluster, the steep spectrum source F, and the relic
area containing filaments G, H, and the tail of source C. The complex
involving sources A and B will not be discussed.

\subsection{Halo flux at 351~MHz}

The largest structure visible in
Fig.~\ref{brentjens_abell2256_fig:a2256_total_intensity} is the diffuse emission
approximately centred on the X-ray source at J2000 position
17$^\mathrm{h}$4$^\mathrm{m}$2$^\mathrm{s}$
+78\degr37\arcmin55\arcsec\ \citep{EbelingEtAl1998}, and extending to
the sources F, O, L, H, and G. This halo source has been described by
several authors
{\citep{BridleFomalont1976,BridleEtAl1979,Kim1999,ClarkeEnsslin2006}.
It is considered to be the main contributor to the total flux of the
cluster at decametric wavelengths \citep{CostainEtAl1972}.

The total flux of the halo at 351~MHz was estimated using the data
selection from
Fig.~\ref{brentjens_abell2256_fig:a2256_total_intensity}. The flux
estimate is complicated by several bright, extended radio sources in
the cluster, which have to be subtracted first. Because of their
complex nature, the sources were subtracted in the image plane. Two
different approaches were used to estimate the halo flux in the area
enclosed by the thick, grey contour in
Fig.~\ref{brentjens_abell2256_fig:a2256_halo}.

In the first approach the CLEAN model was set to zero inside the areas
enclosed by the white contours in
Fig.~\ref{brentjens_abell2256_fig:a2256_halo}. The model was
subsequently convolved with a circular Gaussian of $67\arcsec$ FWHM and
added to the residual image. \modified{Let $\Phi$ be the set of $n_\mathrm{\Phi}$
pixels enclosed by the grey contour and let $\mu$ be the set of
$n_\mathrm{\mu}$ pixels enclosed by both the grey contour and a white
contour. The integrated halo flux is then the sum of all pixels in $\Phi$
that are not in $\mu$, multiplied by $\frac{n_\mathrm{\Phi}}{n_\mathrm{\Phi} -
n_\mathrm{\mu}}$ and divided by the volume of the $67\arcsec$ circular
Gaussian, which results in a flux of $696\pm8$~mJy.} This is most likely a
low estimate, because the sources A, B, C, D, F, and the
eastern parts of G  and H, are situated predominantly in the brighter
parts of the halo. The fainter south-eastern part of the halo
therefore has a relatively large influence on the sum. 

In the second approach, the CLEAN  model pixels \modified{in $\mu$ were} not set to
zero, but were replaced by a radial basis function interpolation
\citep{CarrEtAl2001} of the pixels that constitute the border of the
areas enclosed by the white contours. This interpolated model was
convolved with a circular Gaussian of $67\arcsec$ FWHM and added to
the residual image. The result is shown in
Fig.~\ref{brentjens_abell2256_fig:a2256_halo}.  The interpolation
works extremely well for sources up to the size of source F. The
interpolation of the vast area of sources A, B, C, G, and H is less
successful and appears a bit too faint in the area of A, B, and C, and
a bit too \modified{bright} in the area of G and H. The total flux was computed by
adding the values of all pixels of the restored image that are \modified{in $\Phi$
and} dividing by the volume of the $67\arcsec$ circular Gaussian. The
result of $831\pm10$~mJy is probably a somewhat high estimate because
of the relatively high interpolated surface brightness of the relic
area, which lacks the filamentary structure seen in the rest of the
halo.  The total halo flux is therefore estimated at $0.76\pm0.07$~Jy,
which is the average of the two approaches. The uncertainty is mainly
systematic.

\begin{figure}
\resizebox{\hsize}{!}{\includegraphics{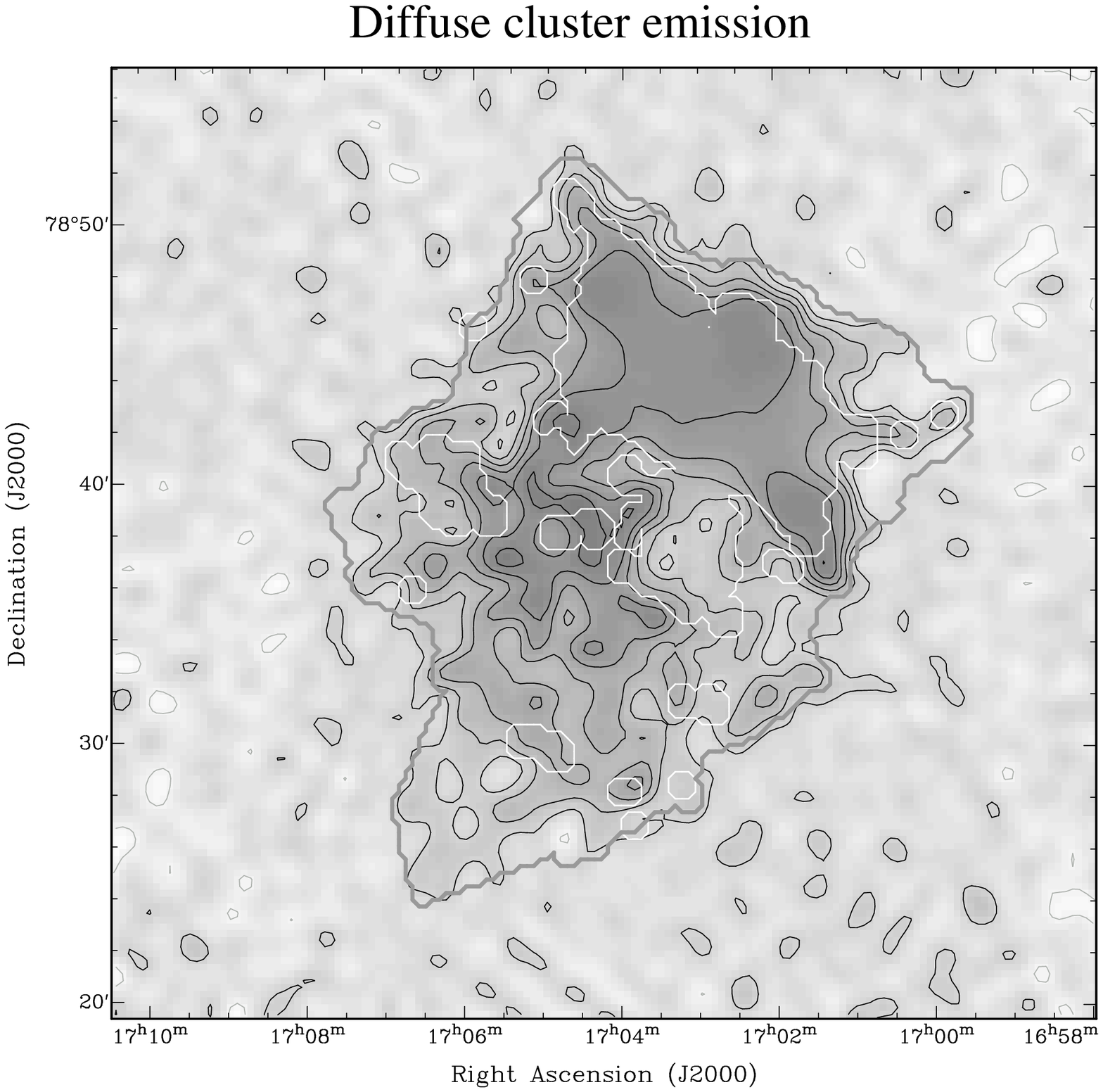}}
\caption{The radio emission of the halo of \object{Abell~2256} at
351~MHz. The contours are drawn at 1, 2, 3, 4, 5, and
6~mJy~beam$^{-1}$ of $67\arcsec$ FWHM. In the areas enclosed by the white
contours, the surface brightness was replaced by a radial
basis function interpolation of the fluxes along the white contours.}
\label{brentjens_abell2256_fig:a2256_halo}
\end{figure}

\subsection{Radio spectrum}

The integrated radio flux of the entire cluster (halo, relic, and
discrete sources combined) was determined from
Fig.~\ref{brentjens_abell2256_fig:a2256_total_intensity} by measuring
the integrated flux within a circle centred at the optical centre of
Abell~2256 at J2000.0 position $\alpha =
17^\mathrm{h}03^\mathrm{m}45^\mathrm{s}$, $\delta =
+78\degr43\arcmin00\arcsec$. The radius of the circle was determined
using Fig.~\ref{brentjens_abell2256_fig:totalflux351}. The derivative
of the integrated flux within a circle with respect to the radius of
the circle is fairly high at small radii, where there is significant
cluster emission. At radii $> 5\arcmin$, the derivative drops sharply
and settles at a more or less constant value at a radius of
$10\arcmin$, which was therefore assumed to be the cluster radius. The
integrated flux within $10\arcmin$ of the optical cluster centre is
$3.51\pm0.06$~Jy at a frequency of 351~MHz.

Table~\ref{brentjens_abell2256_tab:abell2256_flux} lists flux
estimates for the halo as well as the entire cluster including relics,
head-tail galaxies, and background sources.  All fluxes have been
converted to the flux scale of \citet{PerleyTaylor1999} using the flux
of \object{Cas A}, taking its secular variation into account. At
frequencies above 408~MHz, this flux scale is equivalent to the
\citet{BaarsEtAl1977} flux scale.

Note that the flux at 81.5~MHz is twice the flux from
\citet{Branson1967}, as suggested by \citet{MassonMayer1978}. The flux
measurements from \citet{Kim1999} have been read off Fig.~3 of that
paper. The two measurements of the total cluster flux at 408~MHz and
1420~MHz appear to be systematically lower than all other cluster flux
observations.  It was therefore decided not to use these \modified{fluxes} to
determine the average spectra of Abell~2256 (halo plus relic
and discrete sources) and its halo.

The fluxes from Table~\ref{brentjens_abell2256_tab:abell2256_flux} are
plotted in Fig.~\ref{brentjens_abell2256_fig:abell2256_spectrum}.
It is assumed that the cluster flux can be modelled as
the sum of two spectral components,
each with a constant spectral index between 22.25~MHz and 2695~MHz:
\begin{equation}
S(\nu) = S_\mathrm{H,100}\left(\frac{\nu}{100\
\mathrm{MHz}}\right)^{\alpha_\mathrm{H}} +
S_\mathrm{R,100}\left(\frac{\nu}{100\
\mathrm{MHz}}\right)^{\alpha_\mathrm{R}},
\label{brentjens_abell2256_eqn:cluster_spectrum}
\end{equation}
where the first term is due to the halo and the second term is due to
all other sources (relic and discrete sources).
Fitting Eq.~(\ref{brentjens_abell2256_eqn:cluster_spectrum}) to the
total cluster fluxes, assuming uncertainties of 10\% for all flux
points with unknown errors, yields
\begin{equation}
\label{brentjens_abell2256_eqn:haloflux_clusterpoints}
\begin{array}{lcrcl}
S_\mathrm{H,100} & = & 6.0 &\pm &2.1\ \mathrm{Jy}\\
\alpha_\mathrm{H}& = & -1.65&\pm& 0.22\\
S_\mathrm{R,100} & = & 6.8& \pm& 1.8\ \mathrm{Jy}\\
\alpha_\mathrm{R}& = & -0.72&\pm& 0.07.
\end{array}
\end{equation} 
This suggests that the spectrum of the halo may be less steep than the
estimates  by \citet{CostainEtAl1972} ($-1.9$ between 22.25~MHz and
81.5~MHz) and   \citet{BridleEtAl1979} ($-1.8$ between 151~MHz and
610~MHz).

As Fig.~\ref{brentjens_abell2256_fig:abell2256_spectrum} shows, the
halo spectrum lies much closer to the halo fluxes derived in this
paper and by \citet{ClarkeEnsslin2006} than to the flux estimates by
\citet{BridleEtAl1979} and \citet{Kim1999}. \modified{These sets of estimates
are in fact mutually inconsistent.}  The reason is that the maps in
this paper and the maps in \citet{ClarkeEnsslin2006} are much deeper
and detect the halo source over a larger area than
\citet{BridleEtAl1979} and \citet{Kim1999}.  The halo appears to be
roughly circular with a diameter of 12\farcm2 at 1369~MHz
\citep{ClarkeEnsslin2006}.  The map in
Fig.~\ref{brentjens_abell2256_fig:a2256_halo} shows that the halo is
approximately square with a triangular south-eastern extension at
351~MHz. The sides of the square are $18\arcmin$ long at
1~mJy~beam$^{-1}$ and $13\arcmin$ at 3~mJy~beam$^{-1}$. The sides of
the triangular extension are $6\arcmin$ long at 1~mJy~beam$^{-1}$.
The halo is only marginally detected in the 610~MHz map of
\citet{BridleEtAl1979} and at the noise level in their 1415~MHz map.
They estimated its diameter at $9\arcmin$ to $10\arcmin$.
\citet{Kim1999} estimates the halo size slightly larger
($13\arcmin\times10\arcmin$), but the halo is still only marginally
above the noise in the 1420~MHz map presented in that paper. It is
therefore not surprising that the halo flux estimates in these papers
were relatively low.

Using the 1369~MHz halo flux of \citet{ClarkeEnsslin2006}
and the 351~MHz flux derived above, the spectral index of the halo is
estimated at $-1.5\pm0.2$ between these frequencies.
Including the 351~MHz and 1369~MHz halo fluxes in a joint fit with the
total cluster flux points gives much better constrained values between
22.25~MHz and 2695~MHz:
\begin{equation}
\label{brentjens_abell2256_eqn:haloflux_allpoints}
\begin{array}{lcrcl}
S_\mathrm{H,100} & = & 5.9  &\pm& 0.3\ \mathrm{Jy}\\
\alpha_\mathrm{H}& = & -1.61&\pm& 0.04\\
S_\mathrm{R,100} & = & 6.7  &\pm& 0.2\ \mathrm{Jy}\\
\alpha_\mathrm{R}& = & -0.72&\pm& 0.02.
\end{array}
\end{equation} 
The reduced $\chi^2$ of the fit is 0.23, indicating that at least
some fluxes in Table~\ref{brentjens_abell2256_tab:abell2256_flux} are more
accurate than stated.

\begin{figure}
\centering
\resizebox{\hsize}{!}{\includegraphics{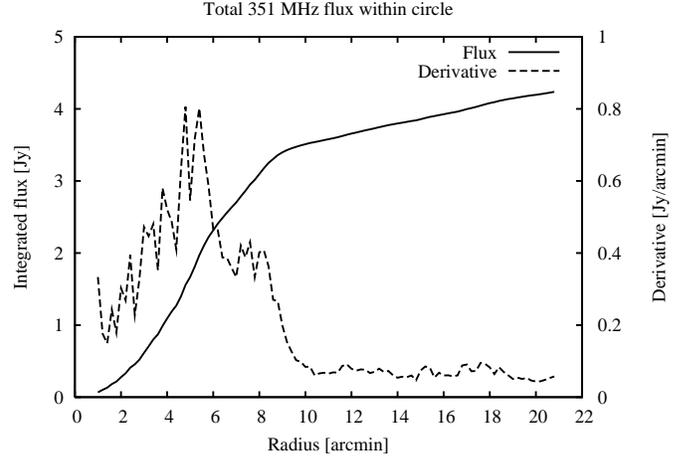}}
\caption{\modified{The solid line is the total 351~MHz radio flux in Jy within a
  circle centred on the optical cluster centre as a function of the
  radius of the circle in arcminutes. The dotted line is the
  derivative with respect to the radius.}}
\label{brentjens_abell2256_fig:totalflux351}
\end{figure}

\begin{table}
\caption{\modified{Flux measurements for Abell~2256.}}
\label{brentjens_abell2256_tab:abell2256_flux}
\begin{minipage}{\columnwidth}
\begin{center}
\begin{tabular}{l@{\ \ }l@{ $\pm$ }l@{ }l@{\ \ }l@{\ \ }l}
  \hline
  \hline
  $\nu$ 
  & \multicolumn{2}{l}{$S$\footnote{\modified{All values have been converted to
      the \citet{PerleyTaylor1999} flux scale using Cas~A.
      The error  is omitted if no reliable error estimate is known.}}} 
  & C/H\footnote{\modified{Indicates whether the flux refers to the entire
    cluster (C) or the halo only (H).}}
  & Fit\footnote{\modified{A "$-$" means that the point is for some reason \emph{not} used to
    determine the spectrum of the halo and the cluster. These reasons are
    described in the text.}}
  & Reference\footnote{\modified{References:
    (C72) \citet{CostainEtAl1972}; 
    (W66) \citet{WilliamsEtAl1966};
    (B67) \citet{Branson1967};
    (M78) \citet{MassonMayer1978};
    (B08) This paper;
    (K99) \citet{Kim1999};
    (B76) \citet{BridleFomalont1976};
    (O75) \citet{Owen1975};
    (H78) \citet{HaslamEtAl1978};
    (B79) \citet{BridleEtAl1979};
    (C06) \citet{ClarkeEnsslin2006}.}} \\
  (MHz) & \multicolumn{2}{l}{(Jy)} &     &    &          \\
  \hline
  22.25 & 100  &32     &C&+ & C72\\
  38    & 41   & 6     &C&+ & W66\\
  81.5  & 17   & 2     &C&+ & B67/M78\\ 
  151   & 8.1  & 0.8   &C&+ & M78\\
  351   & 3.51 & 0.06  &C&+ & \modified{B08}\\
  408   & 2.4  & 0.24  &C&$-$& K99\\ 
  610   & 2.165& \ldots&C&+ & B76\\
  1410  & 1.13 & \ldots&C&+ & B76\\
  1420  & 0.75 & 0.11  &C&$-$& K99\\
  2695  & 0.57 & 0.16  &C&+ & O75\\
  2695  & 0.666& \ldots&C&+ & H78\\ 
  151   & 1.5  & \ldots&H&$-$& B79\\
  351   & 0.76 & 0.07  &H&+ & \modified{B08}\\  
  610   & 0.1  & \ldots&H&$-$& B79\\
  1369  & 0.103& 0.020 &H&+ & C06\\
  1415  & 0.02 & \ldots&H&$-$& B79\\
  1420  & 0.03 & 0.009 &H&$-$& K99\\
  \hline 
  \hline 
\end{tabular}
\end{center}
\end{minipage}
\end{table}

\begin{figure}
\centering
\resizebox{0.7\hsize}{!}{\includegraphics{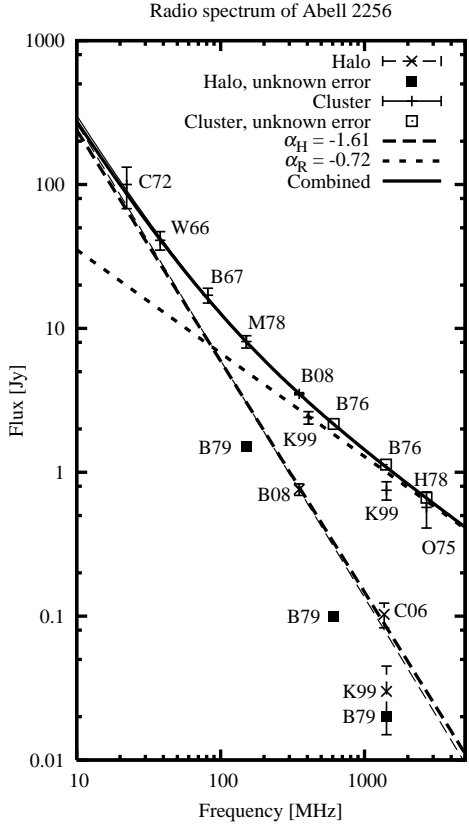}}
\caption{Spectra of the entire cluster and its two spectral components
representing the halo and the rest of the cluster (see
Eq.~(\ref{brentjens_abell2256_eqn:cluster_spectrum})). The labels
indicate the source paper for the data points. Refer to
Table~\ref{brentjens_abell2256_tab:abell2256_flux} for details on the flux
values. The solid lines represent the sum of the two spectral
components as computed by
Eq.~(\ref{brentjens_abell2256_eqn:cluster_spectrum}).
The dashed lines represent the individual spectral components. The
long dashes represent the halo model and the short dashes the
rest of the cluster. The thin curves show the spectral fit of
Eq.~(\ref{brentjens_abell2256_eqn:cluster_spectrum}) to the total cluster flux.
The thick curves show the fit where the \modified{B08} and C06 fluxes of the halo
have been used as extra condition equations on the spectral component
of the halo. The total cluster fluxes of both solutions virtually
overlap.}
\label{brentjens_abell2256_fig:abell2256_spectrum}
\end{figure}

\subsection{Source F}

\begin{figure}
\centering
\resizebox{\hsize}{!}{\includegraphics{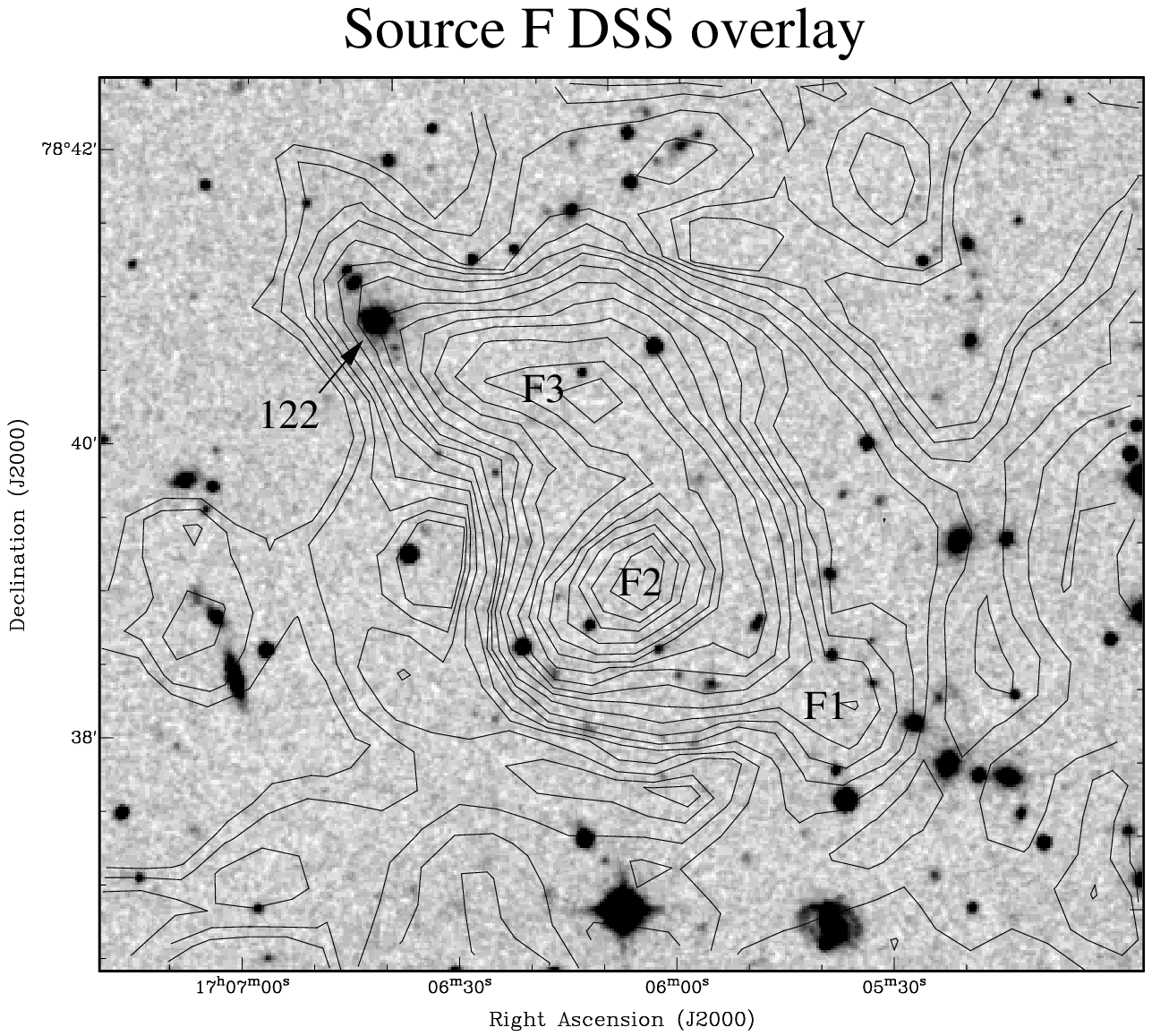}}
\caption{ Palomar DSS blue map of source F overlaid with 351~MHz
Stokes $I$ contours. The contour levels are: 0.2, 0.28, 0.4, 0.56,
\ldots\ mJy~beam$^{-1}$, with a circular Gaussian beam of 
$26\arcsec$ FWHM. The role of Fabricant galaxy 122 is discussed in
the text.}
\label{brentjens_abell2256_fig:source_f_dss}
\end{figure}

\begin{figure}
\centering
\resizebox{0.8\hsize}{!}{\includegraphics{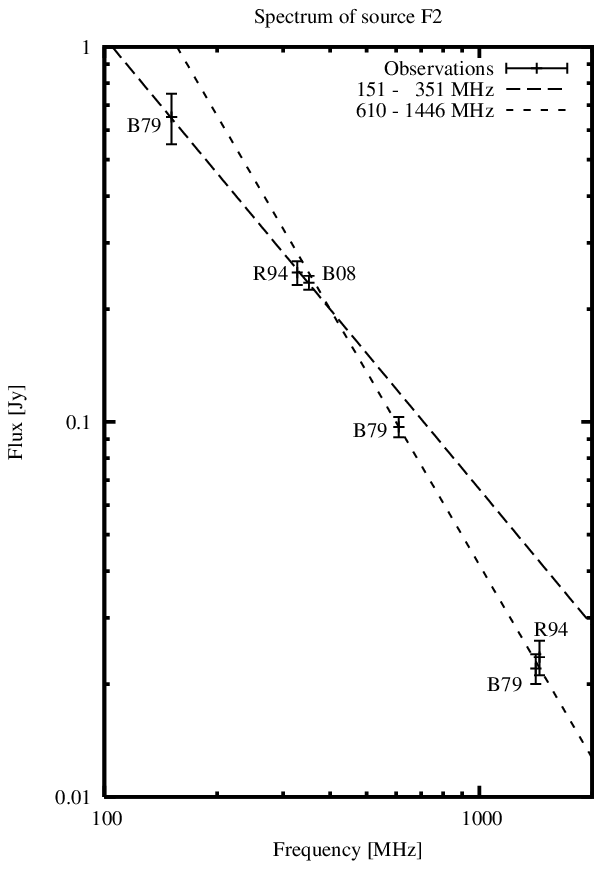}}
\caption{\modified{Radio spectrum of source F2. The labels of the data points
  indicate its source (see
  Table~\ref{brentjens_abell2256_tab:abell2256_flux}). The error bar
  on the 151~MHz point indicates the uncertainty in the extrapolation
  of the 351~MHz flux with the 338/365~MHz spectral index to 151~MHz
  rather than the uncertainty in the estimate by
  \citet{BridleEtAl1979}.}}
\label{brentjens_abell2256_fig:f2spectrum}
\end{figure}

Source F is a bright, ultra steep spectrum source
\citep{MassonMayer1978,BridleEtAl1979,RottgeringEtAl1994}. An optical
image from the Palomar Digital Sky Survey, overlaid with the Stokes
$I$ contours at 351~MHz is shown in
Fig.~\ref{brentjens_abell2256_fig:source_f_dss}. Source F has three
components: F1 (southwest), F2 (central), and F3 (northeast).
\citet{BridleEtAl1979} suggested that because the spectral index
between 1415~MHz and 610~MHz of the entire source is rather constant,
F1, F2, and F3 could be physically related. They suggested that the
entire source is the radio tail of \citet{FabricantEtAl1989} galaxy
122 at the northeastern tip of F3, and that F2 is a section of the
tail oriented exactly along the line of sight. This would explain the
shell-like structure observed in \citet{RottgeringEtAl1994} and
\citet{MillerOwenHill2003}.  Based on these maps, the size of F2 is
estimated at $45\arcsec\times90\arcsec \approx 50\times100$~kpc.
After subtracting Fig.~\ref{brentjens_abell2256_fig:a2256_halo} from
Fig.~\ref{brentjens_abell2256_fig:a2256_total_intensity} in order to
remove the contribution of the halo, the total flux of source F at
351~MHz is $375\pm15$~mJy. The 351~MHz flux of component F2 is
$235\pm10$~mJy. The peak brightness of F2 is 185~mJy~beam$^{-1}$ of
$67\arcsec$ FWHM.

Figure~\ref{brentjens_abell2256_fig:f2spectrum} shows the spectrum of
component F2. The following data are included: $0.65$~Jy at 151~MHz,
$250\pm18$~mJy at 327~MHz, $235\pm10$~mJy at 351~MHz, $97\pm6$~mJy at
610~MHz, $22\pm2$~mJy at 1415~MHz, and $23.6\pm2.5$~mJy at 1446~MHz.
Note that \citet{BridleEtAl1979} do not mention an error for their
151~MHz estimate.  The spectrum appears to be considerably steeper at
frequencies above 351~MHz than it is below that frequency, hence two
power laws of the form
\begin{equation}
S(\nu) = S_{100}\left(\frac{\nu}{100\ \mbox{MHz}}\right)^\alpha
\end{equation}
were fit to the data. Between 151~MHz and 351~MHz
$S_{100}=1.06\pm0.07$ and $\alpha=-1.20\pm0.05$. Between 610~MHz and
1446~MHz $S_{100}=2.15\pm0.40$ and $\alpha=-1.71\pm0.08$. Based on the
data in Fig.~\ref{brentjens_abell2256_fig:spectralindexmap} it was
determined that the spectral index of F2 between 338~MHz and 365~MHz
is $-1.2\pm 0.2$, which is consistent with the low frequency fit in
Fig.~\ref{brentjens_abell2256_fig:f2spectrum}. Extrapolating the
351~MHz flux to 151~MHz using the 338/365~MHz spectral index yields
$0.65\pm0.1$~Jy, which is equal to the estimate by
\citet{BridleEtAl1979}.

The magnetic field in F2 was estimated using the minimum energy
formula given by \citet{BeckKrause2005}. Unfortunately, the steepening
in the spectrum complicates the estimate. Approximating a convex
spectrum with a single power law that is tangent to the spectrum at
some point will overestimate the strength of the magnetic field
because the synchrotron intensity, and therefore the cosmic ray energy
density, is overestimated at both lower and higher energies than the
frequency at which the power law is tangent. \modified{Because the flux is less
affected by synchrotron losses at lower frequencies, the 151~MHz flux
estimate by \citet{BridleEtAl1979} was used, yielding an average
brightness of $0.17\pm0.08$~Jy~arcmin$^{-2}$ if a Gaussian shape of
$1.7\pm0.2\arcmin \times 2\pm0.2\arcmin$~FWHM is assumed. An injection
spectrum of $\alpha \approx -0.7$ was adopted.}

Based on its appearance at 1.4~GHz
\citep{RottgeringEtAl1994,MillerOwenHill2003,ClarkeEnsslin2006} the
most likely morphologies are a spherical shell or a tube viewed along
its axis.  The line of sight through the source is approximately
100~kpc in case of a shell and a few hundred kpc in case of a tube.
Given that the tail of a typical head-tail galaxy is somewhere between
100~kpc and 1~Mpc long, an estimated line of sight of 500~kpc appears
reasonable.  \modified{The magnetic field in source F2 was assumed to be
tangled, which seems valid given its low polarization fraction at
1.4~GHz \citep{ClarkeEnsslin2006}. The minimum energy magnetic field
in source F2 is estimated to be
$2.5\pm0.3\times(K_0+1)^{1/(\alpha+3)}$~$\mu$G for a line of sight of
100~kpc and at most $1.6\pm0.2\times(K_0+1)^{1/(\alpha+3)}$~$\mu$G for
a line of sight of 500~kpc.}

\modified{
According to \citet{BeckKrause2005}, a number density ratio of protons
and electrons of $K_0 = 100$ appears reasonable for several
acceleration mechanisms, including Fermi shock acceleration, secondary
electron acceleration, and plasma turbulence. Note that $K_0$ is not
the same as $\mathcal{K}$, the \emph{energy} density ratio of
relativistic protons and electrons, which is traditionally assumed to
be 1 in a radio lobe plasma. Assuming $K_0=100$ gives
$7.3\pm1.1$~$\mu$G and $4.8\pm0.8$~$\mu$G for lines of sight of
100~kpc and 500~kpc, respectively. These are the values used in the
subsequent synchrotron age estimate.}

It is possible to estimate the age of a radio source based on the
shape of its synchrotron spectrum. The shape of the spectrum reflects
the history of the energy distribution of the injected relativistic
electrons and the effect of various energy loss mechanisms. In the
following discussion it is assumed that synchrotron emission and
inverse Compton scattering of CMB photons dominate the energy loss,
and that the energy distribution of the injected electrons follows a
power law: $n_\mathrm{e}(E)\ \mathrm{d}E \propto E^{-\gamma}\
\mathrm{d}E$
\citep[e.g.,][]{GinzburgSyrovatskii1965,VanderlaanPerola1969,JaffePerola1973}. It
is furthermore assumed that pitch angle scattering is effective, i.e.,
the momentum vectors of the relativistic electrons are distributed
isotropically at all times.

\begin{figure}
\resizebox{\hsize}{!}{\includegraphics{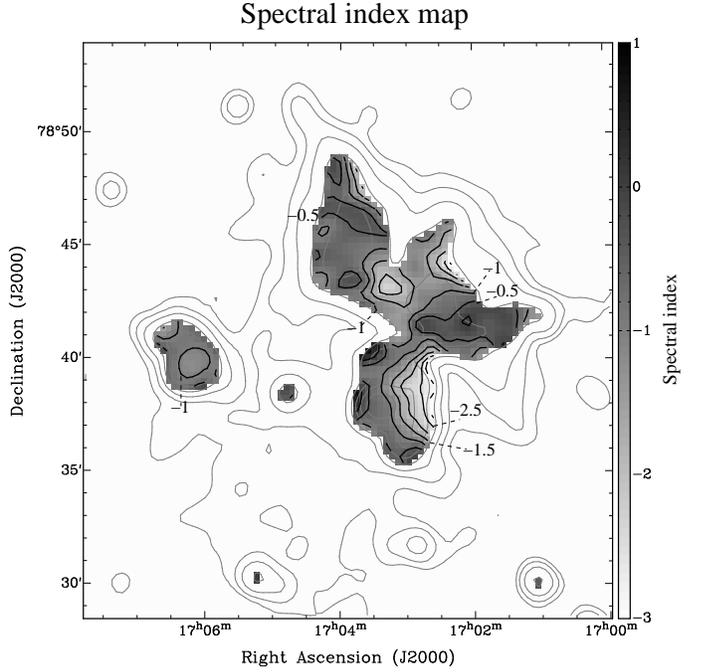}}
\caption{Spectral index map of \object{Abell~2256} between 338~MHz and
365~MHz. The Stokes $I$
contours are drawn at 5, 10, 20, 40, 80, 160, 320 $\sigma$ of the
338~MHz map.  The spectral index contours are drawn at $-3$--$+1$ in
steps of 0.5. The accuracy is $0.46$ at the 40 sigma contour and
$0.23$ at the 80 sigma contour.}
\label{brentjens_abell2256_fig:spectralindexmap}
\end{figure}

\modified{In the simplest scenario}, the energy spectral index $\gamma$ and total
power of  the injected electrons is constant. In this case, the radio
spectrum consists of a power law with spectral index
$\alpha_\mathrm{low} = -(\gamma - 1)/2$ below the break frequency and
a power law with spectral index $\alpha_\mathrm{high} = -\gamma/2$
above the break frequency. The half life time of synchrotron emitting
electrons emitting in a magnetic field with strength $B$ in $\mu$G is 
\begin{equation}
\tau = \frac{2.6\cdot10^{10}\ \mathrm{years}} {B^2 +
B^2_\mathrm{IC}} \sqrt{\frac{B}{(1+z) \nu_\mathrm{b}}}, 
\label{brentjens_abell2256_eqn:classical_synchrotron_age}
\end{equation}
where $\nu_\mathrm{b}$ is the break frequency in MHz and
$B_\mathrm{IC}$ is the strength of a magnetic field with the
same energy density as the CMB.  Assuming a CMB temperature of
2.725~K, its value is $3.238(1+z)^2\ \mu$G.  The power laws fitted to
the data in Fig.~\ref{brentjens_abell2256_fig:f2spectrum} intersect at
$400\pm180$~MHz. Substituting this value for the break frequency and
the minimum energy magnetic fields determined above yields \modified{an age
of $51_{-9}^{+18}$~million years if $B = 7.3$~$\mu$G and
$77_{-11}^{+27}$~million years if $B = 4.8$~$\mu$G.}

The above model is somewhat problematic because it implies that
$\gamma \approx 3.4$, whereas in most head-tail sources $\gamma\approx
2.4$.  It can therefore not be excluded that there is an additional
break in the spectrum below $151$~MHz, where the spectral index
flattens to $\alpha\approx -0.7$. \modified{This situation can occur when the
power of the injection of relativistic plasma has decreased sharply
after a period in which it was high and steady.} The resulting spectrum
has a spectral index of $-0.7$ below the hypothetical break at $\nu <
151$~MHz and a spectral index of $-1.2$ above this break, followed by
an exponential decrease. \modified{This increases the age estimates to $>
83$~million years if $B = 7.3$~$\mu$G and $> 125$~million years $B =
4.8$~$\mu$G. New observations below 150~MHz are needed in order to
establish whether there is indeed an additional break in the
spectrum. It is interesting to estimate the frequency range in which
this break could occur.}

Let us for the moment assume that F2 is indeed part of the tail of
Fabricant galaxy 122. The galaxy is located at a distance of
$2\farcm6$ (180~kpc) from F2 in the plane of the sky. If F2 is indeed
a section of the tail seen on-axis, one would expect that the galaxy
travelled at least this far along the line of sight.  The galaxy must
therefore have travelled at least 0.25~Mpc. The current line-of-sight
velocity of the galaxy is approximately 500~km~s$^{-1}$
\citep{BerringtonLuggerCohn2002}. However, \modified{it is possible} that the
galaxy changed direction between F2 and F3 and is currently travelling
\modified{approximately} in the plane of the sky. Assuming that the actual
velocity is of the order of the velocity dispersion of the cluster of
approximately $1300$~km~s$^{-1}$ \citep{BerringtonLuggerCohn2002}, the
elapsed time since starting the electron injection into source F2 is
at least 200 million years. A magnetic field of the \modified{order of
$7.3$~$\mu$G then gives a break frequency of 26~MHz. This frequency
will decrease if the source is older than 200~million years, but will
increase if the magnetic field is lower than $7.3$~$\mu$G.}
Observations between 151~MHz and 10~MHz will be possible with LOFAR in
the very near future, but will be very difficult below 20~MHz. A
complete spectrum from 10~MHz to 5~GHz will enable a more accurate
minimum energy magnetic field estimate by properly integrating the
cosmic ray energies over the observed spectrum, instead of using a
single power law. \modified{Until} a better constrained synchrotron age for this
source is obtained, its association with galaxy 122 remains uncertain.

\subsection{Relic area}

\begin{figure}
\begin{center}
{\large Relic without compact sources}
\end{center}
\vspace{-3em}
\centering
\resizebox{\hsize}{!}{\includegraphics{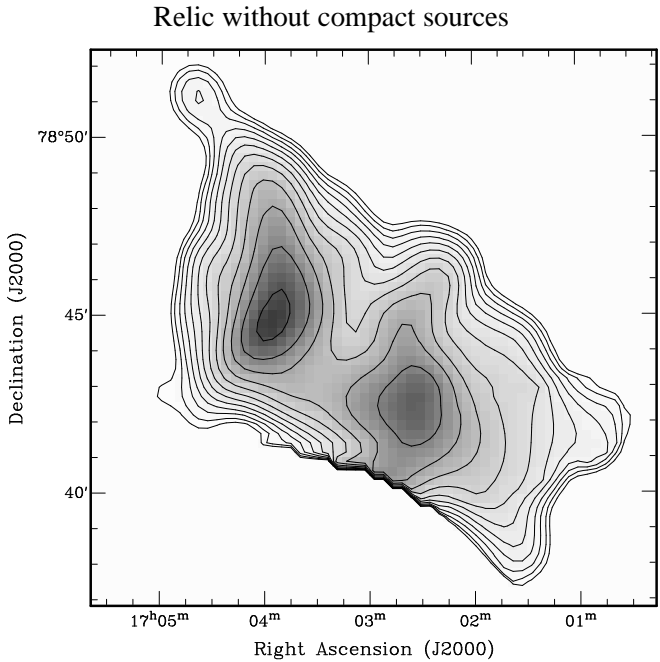}}
\caption{\modified{The relic area after subtraction of the halo and the tail of
  source C. The lowest contour is drawn at $1.5$~mJy~beam$^{-1}$ and
  all subsequent contours are scaled by $\sqrt{2}$. The resolution is
  $67\arcsec$ FWHM.}}
\label{brentjens_abell2256_fig:relic-interpolated}
\end{figure}

The northwestern area of the cluster, dominated by the filaments G, H,
and source C is perhaps the most intriguing part of Abell~2256.
Most structure in that part is blended by the low resolution of
Fig.~\ref{brentjens_abell2256_fig:a2256_total_intensity}, but
Fig.~\ref{brentjens_abell2256_fig:cleanmodel} shows considerable
detail. 

At the resolution of Fig.~\ref{brentjens_abell2256_fig:cleanmodel},
the long, straight tail of source C is striking. It is 11\farcm5\
(780~kpc) long at 351~MHz, which is considerably longer than the
7\farcm08\ (480~kpc) at 1446~MHz \citep{RottgeringEtAl1994}. At
approximately $6\arcmin$ from the host galaxy, the tail begins to
bend: first towards the northwest, then just after source I to the
west-southwest and finally back to the northwest near source S. The
period of the tentative oscillation is approximately 6\arcmin\
(400~kpc) and the amplitude is about 0\farcm6\ (40~kpc).

Although it is not as detailed as the 1369~MHz VLA C configuration
image of \citet{ClarkeEnsslin2006},
Fig.~\ref{brentjens_abell2256_fig:cleanmodel} clearly shows the
filamentary nature of this part of the cluster.  The FWHM of the
brighter parts of filaments G and H is of the order of of
60\arcsec$\pm$15\arcsec\ or 70$\pm$18~kpc. Filament G is approximately
10\arcmin\ (680~kpc) long and filament H is approximately 8\arcmin\
(550~kpc) from the tail of source C to the border of the relic
emission directly south of source AC. There is a long, straight ridge
south of the straight part of the tail  of source C, at a position
angle of approximately $210\degr$\ (N through E). In fact this source
may extend north of the tail of C into the southern part of filament
H. If that is indeed the case, the source is $8\farcm8$ (600~kpc)
long. It appears to be unresolved at $26\arcsec$, limiting its width
to less than 35~kpc.

The flux of the relic area was determined after subtracting
Fig.~\ref{brentjens_abell2256_fig:a2256_halo} from
Fig.~\ref{brentjens_abell2256_fig:a2256_total_intensity}. Sources A,
B, and C were excluded. The extended tail of source C was removed by
interpolating fluxes of pixels north and south of the tail using
radial basis function interpolation \citep{CarrEtAl2001}. The
resulting map is shown in
Fig.~\ref{brentjens_abell2256_fig:relic-interpolated}. \modified{After
correcting for the fact that the interpolated halo brightness in the
relic area in Fig.~\ref{brentjens_abell2256_fig:a2256_halo} appears to
be on average $1.5\pm0.5$~mJy~beam$^{-1}$ too high, the total flux in
the relic area is $1.39\pm0.07$~Jy. This is approximately} a factor of
three larger than the sum of the fluxes of the filaments G and H as
determined by \citet{RottgeringEtAl1994}. This discrepancy can be
explained by the difference in uv coverage between the WSRT at 351~MHz
and the VLA in B configuration at 327~MHz, which is not sensitive to
scales above $7\arcmin$.  The uv-coverage of the VLA-D observation by
\citet{ClarkeEnsslin2006} at 1369~MHz is much more similar to the WSRT
at 351~MHz. Using their estimate of the relic flux, \modified{the spectral index
between 1369~MHz and 351~MHz is $-0.81\pm0.05$.} This is much steeper
than the 1446/327~MHz spectral index by \citet{RottgeringEtAl1994},
but slightly flatter than the average 1703/1369~MHz spectral index that
\citet{ClarkeEnsslin2006} found.

\begin{table*}
\caption{\modified{Rotation measures and linearly polarized fluxes of discrete
sources within 1\fdg 5 of the optical centre of Abell~2256.}}
\label{brentjens_abell2256_tbl:discrete_sources}
\begin{minipage}{\textwidth}
\begin{center}
\begin{tabular}{llllll}
\hline
\hline
Nr.\footnote{\modified{Label of the source in Fig.~\ref{brentjens_abell2256_fig:galactic_rm}.}}
 & \multicolumn{2}{l}{Position\footnote{\modified{Peak of the \emph{polarized} flux.}}} 
 & $\phi$ 
 & $|P_\mathrm{a}|$\footnote{\modified{Apparent polarized flux.}}
 & $|P_\mathrm{i}|$\footnote{\modified{Polarized flux corrected for primary beam
   attenuation.}}\\
    & (J2000 $\alpha$) & (J2000 $\delta$) & (rad~m$^{-2}$) & (mJy~beam$^{-1}$) & (mJy~beam$^{-1}$)\\
\hline
 1 &  $16^\mathrm{h} 47^\mathrm{m} 02\fs 7$ &
      $+79\degr 06\arcmin 16\arcsec$ &   $-27.0\pm0.1$  &10.1 & 15.0\\ 
 2 &  $16^\mathrm{h} 49^\mathrm{m} 07\fs 7$ & 
      $+78\degr 51\arcmin 02\arcsec$ &   $-28.0\pm0.8$  & 1.4 &  1.8\\ 
 3 &  $16^\mathrm{h} 51^\mathrm{m} 45\fs 1$ & 
      $+78\degr 09\arcmin 33\arcsec$ &   $-29.9\pm0.8$  & 1.2 &  1.7\\ 
 4 &  $17^\mathrm{h} 00^\mathrm{m} 30\fs 7$ &
      $+79\degr 30\arcmin 47\arcsec$ &   $-20.8\pm0.2$  & 5.5 &  7.7\\ 
 5 &  $17^\mathrm{h} 00^\mathrm{m} 53\fs 6$ &
      $+79\degr 30\arcmin 32\arcsec$ &   $-22.1\pm0.2$  & 6.1 &  8.4\\ 
 6 &  $17^\mathrm{h} 05^\mathrm{m} 18\fs 4$ &
      $+77\degr 47\arcmin 21\arcsec$ &   $-24.7\pm0.5$  & 2.0 &  3.2\\ 
 7 &  $17^\mathrm{h} 05^\mathrm{m} 55\fs 5$ & 
      $+77\degr 06\arcmin 30\arcsec$ &   $-18.9\pm0.2$  & 6.5 & 26.0\\ 
 8 &  $17^\mathrm{h} 06^\mathrm{m} 14\fs 4$ & 
      $+77\degr 07\arcmin 47\arcsec$ &   $-17.9\pm0.2$  & 7.1 & 29.4\\ 
 9 &  $17^\mathrm{h} 07^\mathrm{m} 17\fs 0$ & 
      $+78\degr 06\arcmin 31\arcsec$ &   $-30.7\pm0.5$  & 2.0 &  2.5\\ 
10 &  $17^\mathrm{h} 18^\mathrm{m} 44\fs 7$ & 
      $+78\degr 39\arcmin 30\arcsec$ &   $-49.1\pm0.3$  & 3.2 &  4.2\\ 
11 &  $17^\mathrm{h} 23^\mathrm{m} 34\fs 7$ & 
      $+78\degr 29\arcmin 18\arcsec$ &   $-49.8\pm0.5$  & 2.0 &  3.4\\ 
\hline
\hline
\end{tabular}
\end{center}
\end{minipage}
\end{table*}

The spectral index map in
Fig.~\ref{brentjens_abell2256_fig:spectralindexmap} shows that the
365/338~MHz spectral index is far from uniform across the relic area.
The average spectral index of filament H is $-1.17\pm 0.03$, which is
comparable to its 1703/1369~MHz spectral index. The standard deviation
of the spectral index distribution of filament H is 0.34.

The spectral index of the brightest part of filament G is $-0.70\pm
0.1$.  Slightly north of that point is an east-west ridge with a
365/338~MHz spectral index of $-0.3$~to~$-0.4$. There are several
small areas with both steeper and flatter spectra. The average
spectral index of filament G is $-0.76\pm 0.03$, which is
significantly flatter than its 1703/1369~MHz spectral index
\citet{ClarkeEnsslin2006}.  The standard deviation of the spectral
index distribution is 0.37.

\section{Linear polarization}

\begin{figure}
\centering
{\large Galactic Faraday rotation}\\
\resizebox{\hsize}{!}{\includegraphics{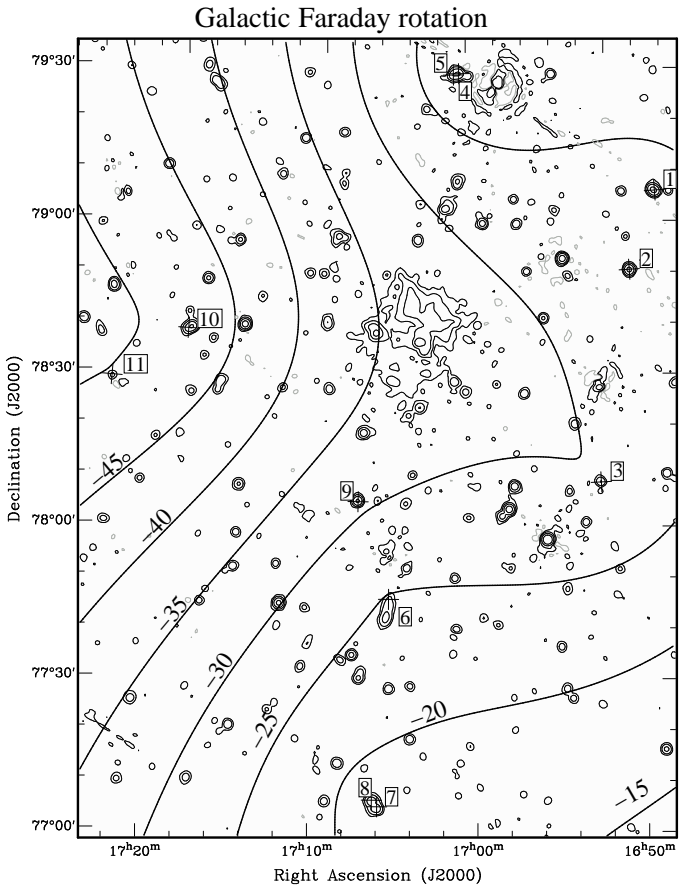}}
\caption{The Galactic Faraday rotation in rad~m$^{-2}$ overlaid on
Stokes $I$ contours. The Galactic RM is drawn at intervals of
$5$~rad~m$^{-2}$ and the total intensity contours at $-3$, 3, 12, 48,
192 $\times$0.6~mJy~beam$^{-1}$. The polarized sources from
Table~\ref{brentjens_abell2256_tbl:discrete_sources} are indicated
with crosses.}
\label{brentjens_abell2256_fig:galactic_rm}
\end{figure}

Although the relic area is between 20\% and 50\% polarized at 1.4~GHz
\citep{ClarkeEnsslin2006,BridleEtAl1979}, there was no linearly
polarized emission at 351~MHz at rotation measures between $-800$ and
$+800$~rad~m$^{-2}$ that could be attributed to the cluster.  The
relic area must therefore be substantially depolarized.  This is not
surprising, as already noted by \citet{Jaegers1987} at 610~MHz. A
search in the area around \object{Abell~2256} for structures similar
to those attributed to  the Perseus cluster
\citep{DeBruynBrentjens2005} or found near \object{Abell~2255} by
\citet{GovoniEtAl2005} did not uncover anything.  The only visible
polarized sources were the Galactic synchrotron foreground at $\phi$
between $-20$ and $-24$~rad~m$^{-2}$ and the collection of nine
discrete sources (two doubles) listed in
Table~\ref{brentjens_abell2256_tbl:discrete_sources}.

These sources made it possible to estimate the Galactic contribution
to the Faraday depth of the relic much more accurately than the
current estimate of $-4\pm37$~rad~m$^{-2}$ \citep{ClarkeEnsslin2006}.
A coarse estimate is obtained by taking the mean value of the Faraday
depths of the sources, which results in $-29\pm3$~rad~m$^{-2}$. 

A better approach is to interpolate the Faraday depths of the sources.
The result is displayed in
Fig.~\ref{brentjens_abell2256_fig:galactic_rm}. The Galactic Faraday
depth field was estimated using a radial basis function interpolation
\citep{CarrEtAl2001}. The interpolated Galactic foreground
contribution in the direction of the relic is $-33\pm2$~rad~m$^{-2}$.
Subtracting this from the average RM of the relic that
\citet{ClarkeEnsslin2006} found ($-44$~rad~m$^{-2}$), yields a cluster
contribution of $-11\pm2$~rad~m$^{-2}$.

\modified{Using an isothermal beta model determined by \citet{MohrEtAl1999} from
ROSAT PSPC data (see Eq.~(\ref{brentjens_abell2256_eqn:beta-model}))
and assuming that magnetic fields are frozen in (see
Eq.~(\ref{brentjens_abell2256_eqn:b_propto_ne})), this corresponds to a
large scale magnetic field strength at the centre of the cluster of at
least 0.08~$\mu$G if the relic were situated at the same distance as
the cluster centre, and at least 0.4~$\mu$G if the relic is 500~kpc
closer to the earth.}

The depolarization of the relic emission gives a handle on the
conditions inside the filaments. Because there is no significant
polarized flux in the relic area at 351~MHz, only upper limits to the
polarization fraction can be established.
Figure~\ref{brentjens_abell2256_fig:polfractionmap} shows a map of the
upper limits to the polarization fraction at 351~MHz. The maximum
value of $|F(\phi)|$ for $-800 \le \phi \le +800$~rad~m$^{-2}$ was
determined for each pixel. Refer to
\citep{Burn1966,GardnerWhiteoak1966,SokoloffEtAl1998,BrentjensDeBruyn2005}
for definitions of the Faraday dispersion function $F(\phi)$.  This
map was then divided by the map in
Fig.~\ref{brentjens_abell2256_fig:a2256_total_intensity}. The
resulting upper limits are approximately a factor of four higher than
upper limits obtained by dividing the image of the RMS value of
$|F(\phi)|$ in the same range of $\phi$ by
Fig.~\ref{brentjens_abell2256_fig:a2256_total_intensity}. Because
$\mbox{RMS} = \sigma\sqrt{2}$ if the noise distribution is Gaussian,
this corresponds approximately to a 6$\sigma$ upper limit.

The fractional polarization is less than 1\% in the brightest parts of
filaments G and H at 351~MHz. The maps by \citet{Jaegers1987} give a
$3\sigma$ upper limit of 20\% at these locations at
608.5~MHz. \citet{ClarkeEnsslin2006} find about 28\% polarization at both
locations  at 1369~MHz, with a resolution of
$52\arcsec\times45\arcsec$. They furthermore find, without precisely
specifying how it is spatially distributed,  that the polarization
fraction at 1703~MHz is generally about 8\% higher than at 1369~MHz,
which would imply roughly 30\% polarization at these locations.

Given the high degree of polarization at a resolution of the order of
an arcminute at 1.4~GHz, the small spatial variance in RM, and the
smooth structure in polarization angle that \citet{ClarkeEnsslin2006}
find, it is unlikely that the sharp decrease in polarization fraction
with increasing wavelength is due to beam depolarization. The narrow
channels, combined with RM-synthesis, rule out bandwidth
depolarization at $|\phi| \lesssim 800$~rad~m$^{-2}$.  This leaves
differential Faraday rotation along the line of sight, caused by the
co-location of emitting plasma and Faraday rotating plasma as the most
likely cause of the depolarization. The depolarization can be used to
determine the Faraday thickness, or the extent in Faraday depth, of
the radio emitting plasma.

\begin{figure}
\centering
{\large 351~MHz fractional polarization limits}\\
\resizebox{\hsize}{!}{\includegraphics{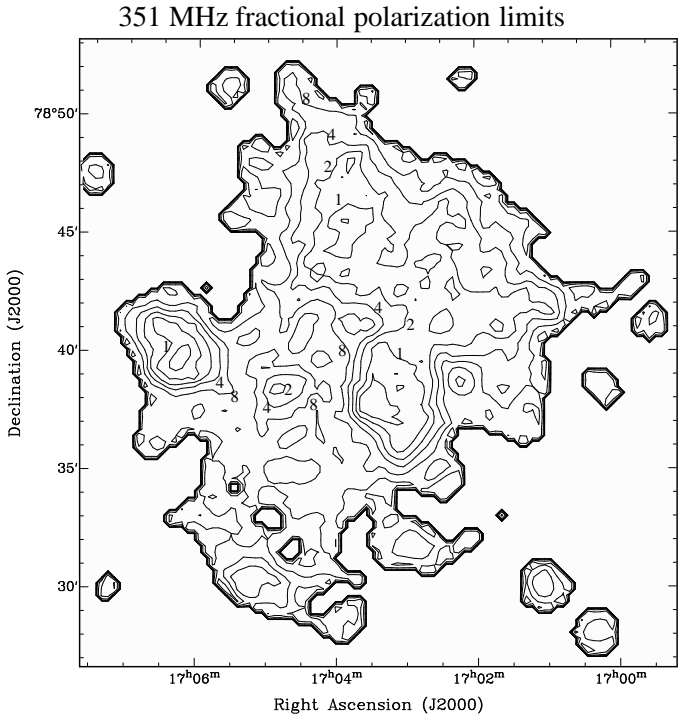}}
\caption{Upper limits ($\approx 6\sigma$, refer to the text for
details) to the polarization fraction in percent at 351~MHz at a
resolution of 67\arcsec\ FWHM. Contours are drawn at 0.5\%, 1\%, 2\%,
4\%, 8\%, 16\%. }
\label{brentjens_abell2256_fig:polfractionmap}
\end{figure}

The depolarization data are plotted in
Fig.~\ref{brentjens_abell2256_fig:depolarization}. Although an
accurate determination of the Faraday thickness requires new,
sensitive polarization observations between \modified{0.4 and 1.4~GHz}, it is
interesting to obtain an estimate of the extent of the relic sources
in Faraday depth using the available detections and limits.

The uniform polarization angle structure at 1.4~GHz indicates that the
magnetic field is relatively regular, perhaps even uniform at scales
of a few arcminutes. If the relic has a uniform electron density, its
structure in Faraday depth can be approximated by a uniform slab model
\citep{Burn1966}. The fractional polarization as a function of
$\lambda^2$ of a uniform slab is a sinc function. The sinc curve in
Fig.~\ref{brentjens_abell2256_fig:depolarization} corresponds to a
slab with a full width of 21~rad~m$^{-2}$ in Faraday depth. As can be
seen in Fig.~\ref{brentjens_abell2256_fig:depolarization}, it is
impossible for a sinc function to simultaneously satisfy the
\citet{ClarkeEnsslin2006} points as well as the upper limits presented
in this paper. The uniform slab model is therefore rejected.  The
Gaussian that is plotted in
Fig.~\ref{brentjens_abell2256_fig:depolarization} has a FWHM of
4.7~rad~m$^{-2}$ in Faraday depth and satisfies all upper limits and
the two points by \citet{ClarkeEnsslin2006}. A Gaussian was chosen
because it is a function that smoothly goes from a high value to a low
value and has an easy analytical Fourier transform. Because this
particular Gaussian satisfies all points rather closely without
exceeding the limits, the FWHM of 4.7~rad~m$^{-2}$ is considered a
lower limit to the extent of filament G in Faraday depth.  A more
physical model will nevertheless have to wait until more sensitive
observations between 400~MHz and 1.4~GHz are available.

\section{Magnetic field in filament G}
\label{brentjens_abell2256_sec:magnetic_field_source_g}

\begin{figure}
\centering
{\large Depolarization of filament G}\\
\resizebox{\hsize}{!}{\includegraphics{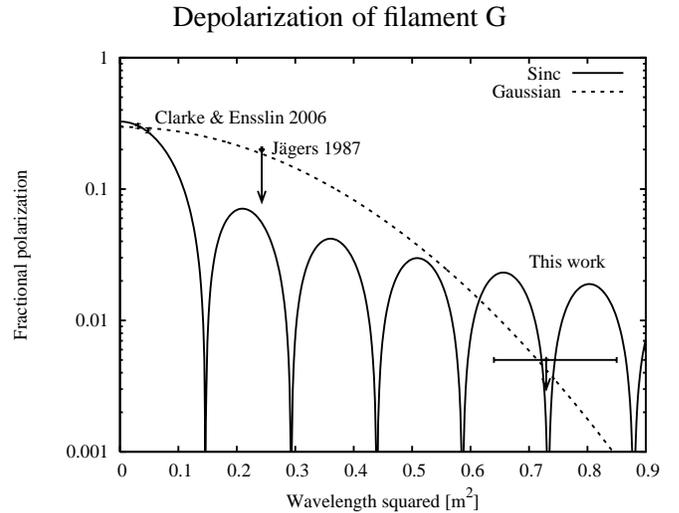}}
\caption{Polarization fraction of the brightest part of filament G as a
function of wavelength squared. The point by \citet{Jaegers1987} is a
3$\sigma$ upper limit. The point at the bottom right is half the limit from
Fig.~\ref{brentjens_abell2256_fig:polfractionmap} and is therefore 
equivalent to a 3$\sigma$ upper limit. The horizontal bars indicate
the $\lambda^2$ coverage of the observations.}
\label{brentjens_abell2256_fig:depolarization}
\end{figure}

\modified{
It is possible, with appropriate assumptions, to derive the
magnetic field strength in filament G from its depolarization
properties as displayed in
Fig.~\ref{brentjens_abell2256_fig:depolarization}.}  

According to the isothermal beta model
\citep{CavaliereFuscoFemiano1976} determined by \citet{MohrEtAl1999}
from ROSAT PSPC data,
\begin{equation}
\frac{n_\mathrm{e}(b,l)}{n_\mathrm{e,0}} = \left(1 + \frac{b^2 +
l^2}{r_\mathrm{c}^2}\right)^{-3\beta/2},
\label{brentjens_abell2256_eqn:beta-model}
\end{equation}
where $n_\mathrm{e}(b,l)$ is the \modified{thermal} electron density at impact
parameter $b$ and line of sight distance between the cluster centre
and the position $l$ (see
Fig.~\ref{brentjens_abell2256_fig:situation}), $n_\mathrm{e,0} =
3.60\pm0.06\times10^{-3}h_{71}^{\frac{1}{2}}$~cm$^{-3}$ is the central
electron density, $r_\mathrm{c} = 342\pm30h_{71}^{-1}$~kpc is the core
radius, and $\beta=0.828\pm0.06$ is the ratio of the specific kinetic
energies of galaxies and gas.  The impact parameter of the brightest
part of filament G with respect to the centre of the X-ray source is
6\farcm9 (470$\pm 15$~kpc). \modified{\citet{ClarkeEnsslin2006} established that
the relic must be on the front side of the cluster due to the small
variance of the observed RMs. In the remainder of this section it is
assumed that the brightest point of filament G lies approximately
500~kpc in front of the centre of the cluster ($l\approx
-500$~kpc). The expected thermal electron density in the neighbourhood
of the filament is then approximately $4.8\times10^{-4}$~cm$^{-3}$.
}

Following the reasoning of \citet{MurgiaEtAl2004}, the magnetic field
strength is assumed to be a power law of the thermal electron density:
\begin{equation}
\|\vec{B}\| = a n_\mathrm{e,th}^{\mu/\beta},
\label{brentjens_abell2256_eqn:b_propto_ne}
\end{equation}
where $a$ is the proportionality constant.  If the magnetic energy
density is proportional to the gas energy density, $\mu = \beta/2$. If
the magnetic field is frozen in the plasma, then $\mu = 2\beta/3$. I
only consider the latter case because the difference between the two is
barely noticeable in view of the uncertainties involved.

\begin{figure}
\begin{center}
{\large Sketch of the cluster}\\
\vspace{0.5em}
\resizebox{0.7\hsize}{!}{\includegraphics{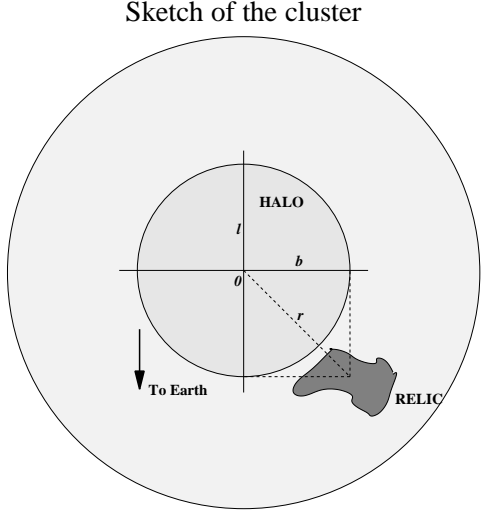}}
\end{center}
\caption{\modified{Coordinate system used for the computations in
  Sect.~\ref{brentjens_abell2256_sec:magnetic_field_source_g}.}}
\label{brentjens_abell2256_fig:situation}
\end{figure}

\modified{
The relativistic electrons in relic sources such as filaments G and H
are assumed to be shock accelerated \citep{EnsslinEtAl1998}. Merger
shocks and accretion shocks typically compress the gas by a factor of
up to 4 \citep{BlandfordOstriker1978}. For the sake of simplicity it
was assumed that the resulting thermal electron density as a function
of position along the line of sight has an offset Gaussian shape:
\begin{equation}
  \label{brentjens_abell2256_eqn:ne_thermal}
  n_\mathrm{e,th}(\vec{l} -\vec{l_0}) = n_\mathrm{e,0}\left(
    1 + \left(r - 1\right)\mathrm{e}^{-4\log{2}\left(\frac{\vec{l}-\vec{l_0}}{w}\right)^2}
  \right),
\end{equation}
where $r$ is the shock compression ratio, $w$ is the FWHM of the
compressed area, and $n_\mathrm{e,0} = 4.8\times10^{-4}$~cm$^{-3}$ is
the expected electron density near the relic as calculated using
Eq.~(\ref{brentjens_abell2256_eqn:beta-model}).}

\modified{
The synchrotron radiation is caused by relativistic electrons, which
have a much lower density. The synchrotron intensity from a slice of
plasma with infinitesimal thickness $\mathrm{d}x$ is
\citep[see e.g.][]{PfrommerEnsslinSpringel2008-II}
\begin{equation}
  \label{brentjens_abell2256_eqn:synchrotron_emissivity}
  I(l)\ \mathrm{d}l \propto \frac{n_\mathrm{e,rel}\|\vec{B}\|^{1-\alpha}}
  {\|\vec{B}\|^2 + B_\mathrm{IC}^2}\ \mathrm{d}l,
\end{equation}
where $B_\mathrm{IC} = 3.238(1+z)^2\ \mu$G is the contribution from
inverse Compton losses due to cosmic microwave background photons and
$\|\vec{B}\|$ is given by
Eq.~(\ref{brentjens_abell2256_eqn:b_propto_ne}).}

\modified{
It is assumed that there are only significant amounts of relativistic
electrons in the compressed area. More specifically, the relativistic
electron density is assumed to be a Gaussian with unit peak and the
same FWHM $w$ as the compression peak in the thermal electron
distribution. The absolute number of relativistic electrons is
unimportant for depolarization arguments. Only the shape of the
distribution
matters. Figure~\ref{brentjens_abell2256_fig:model-electron-density}
shows an example of a thermal electron density profile with its
accompanying relativistic electron density profile. Note that the
vertical scale of the relativistic profile is arbitrary. More precise
modelling is not useful at this point because there are only upper
limits available below 1369~MHz.}

\modified{
The Faraday depth $\phi(l)$ of a given location is obtained by
integrating
\begin{equation}
  \frac{\mathrm{d}\phi}{\mathrm{d}l} = 0.81 a\times n_\mathrm{e,th}^{\frac{5}{3}}
  \label{brentjens_abell2256_eqn:phi-of-x}
\end{equation}
from $l = -2w$ to the location of interest. The Faraday dispersion
function
\begin{equation}
  F(\phi) =  I(\phi)/\frac{\mathrm{d}\phi}{\mathrm{d}l}.
  \label{brentjens_abell2256_eqn:f-phi-relic-g}
\end{equation}
}

\begin{figure}
  \resizebox{\hsize}{!}{\includegraphics{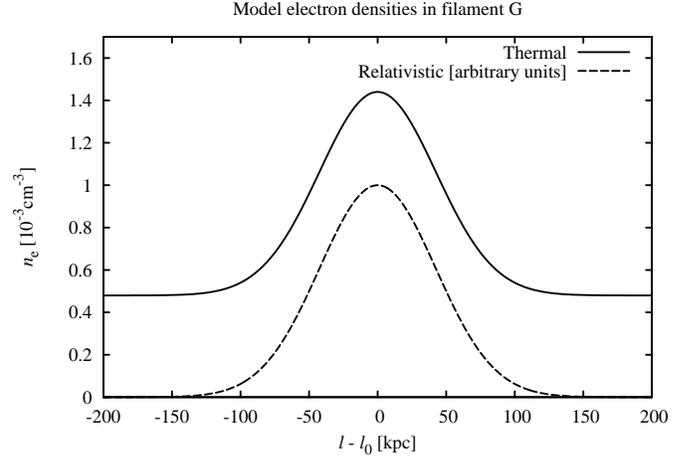}}  
  \caption{\modified{Model electron densities used for estimating the
    magnetic field in filament G. The model densities are plotted as a
    function of position along the line of sight with respect to the
    centre of filament G. The FWHM of the peaks is 100~kpc. The
    vertical scale applies to the thermal electron density only. The
    peak gas compression ratio in this plot is three.}}
  \label{brentjens_abell2256_fig:model-electron-density}
\end{figure}

\modified{
From symmetry arguments one expects the FWHM thickness of the emitting
filaments in the relic to be of the order of
30--100~kpc. Figure~\ref{brentjens_abell2256_fig:b-field-filament-g}
shows the peak magnetic field in the relic as a function of the shock
compression ratio for FWHM thicknesses from 25 to 400~kpc. The peak
field strength was derived by finding the value of $a$ that created a
FWHM of $F(\phi)$ of either 21 or 4.7~rad~m$^{-2}$, which was
subsequently substituted into
Eq.~(\ref{brentjens_abell2256_eqn:b_propto_ne}).}

\modified{
As Fig.~\ref{brentjens_abell2256_fig:b-field-filament-g} shows, the
field estimate depends on the assumptions about the shock compression
ratio and the extent of the filament in both geometric space as well
as Faraday space. A value of the order of $0.2$~$\mu$G appears
reasonable assuming a relic thickness of the order of 30~kpc and an
extent in Faraday depth of 4.7~rad~m$^{-2}$ FWHM. However, due to the
uncertainties involved this value may be off by a factor of up to 10.
}


\section{Conclusions and future work}

The spectral index of the diffuse central halo of \object{Abell~2256}
is $-1.61\pm0.04$. \modified{Although this value is steeper than the
  610/1415~MHz value of $-1.2$ in \citet{BridleEtAl1979}, it is
  somewhat flatter than their 150/610~MHz spectral index ($-1.8$) and
  considerably flatter than the value of $-2.04$ derived by
  \citet{Kim1999}.} The reason for this is twofold:
\begin{itemize}
\item in \citet{ClarkeEnsslin2006} and in this work, the halo was
  detected over a larger extent than in previous observations by
  \citet{BridleEtAl1979} and \citet{Kim1999} due to the increased
  sensitivity of the observations;
\item \citet{Kim1999} did not correct the 81.5~MHz flux
  \citep{Branson1967} as proposed by \citet{MassonMayer1978}.
\end{itemize}

The \modified{observed spectral index} puts the classical minimum energy magnetic
field in the range of 2--4~$\mu$G, and the hadronic minimum energy
field between 4 and 9~$\mu$G \citep{ClarkeEnsslin2006}.

Despite the deep polarization images obtained after RM-synthesis, no
polarized emission was detected that could be attributed to
Abell~2256. The complete depolarization of filaments G and H
was expected because of the relatively high density environment in
which they reside. The upper limit to the fractional polarization of
less than 1\% is consistent with the 608.5~MHz upper limit by
\citet{Jaegers1987} and the depolarization between 1705~MHz and
1395~MHz that \citet{ClarkeEnsslin2006} found.

\begin{figure}
\resizebox{\hsize}{!}{\includegraphics{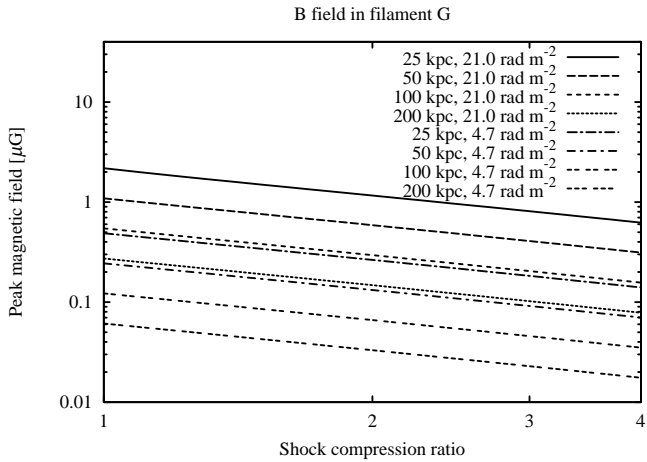}}
\caption{\modified{Estimates of the magnetic field parallel to the line of sight
  in filament G.}}
\label{brentjens_abell2256_fig:b-field-filament-g}
\end{figure}

\modified{
Reliable measurements of the magnetic field in relic sources are
relatively rare. \citet{BagchiPislarLimaNeto1998} used inverse Compton
scattering to derive the cluster scale magnetic field in the relic of
\object{Abell~85}. They found a value of
$0.95\pm0.10$~$\mu$G. \citet{ChenHarrisHarrisonMao2008} derived
classical equipartition field strengths of 0.63 and 1.3~$\mu$G for the
relics in \object{0917+75} and \object{1401-33} respectively.  Their
$3\sigma$ inverse Compton lower limits are 0.81 and 2.2~$\mu$G.
\citet{EnsslinEtAl1998} listed an equipartition field of $3
h_{71}^{2/7}$~$\mu$G for the entire relic in
Abell~2256. Although the estimate of the line-of-sight field
in filament G presented in this work is not very accurate
($0.02$--$2$~$\mu$G with $0.2$~$\mu$G in case of reasonable
assumptions about the shock compression ratio and path length along the
line of sight), it is consistent with the quoted literature values for
magnetic fields in relic sources in general. An actual measurement of
the depolarization of filament G can improve the estimate
significantly, although the major uncertainty is in the physical
thickness of the relic along the line of sight, which is very
difficult to obtain.}

Polarization observations covering frequencies between 1.4~GHz and
400~MHz, where the fractional polarization decreases rapidly, would
make it possible to recover the (linearly polarized) emissivity of
filaments G and H as a function of Faraday depth along the line of
sight.  The high fractional polarization at 1.4~GHz ensures that the
magnetic field has no reversals along the line of sight. Therefore,
the Faraday depth of a certain location in the relic sources should be
a monotonic function of the geometric distance between the front of
the sources and that location, enabling the first 3D reconstruction of
an extragalactic synchrotron source.

The fact that no evidence was found for polarized emission from LSS
shocks or buoyant bubbles further away from the cluster centre than
filaments G and H could have several reasons:
\begin{itemize}
\item they are depolarized due to internal Faraday dispersion, just
like sources G and H;
\item there are no shocks oriented edge-on, only face-on, resulting in
an undetectably low surface brightness;
\item apart from the relic area, there are simply no other bubbles or
large scale shocks in Abell~2256 that emit radio waves.
\end{itemize}

The first point can be tested by applying RM-synthesis to deep
polarimetry at 1.4~GHz. Discriminating between the other two
possibilities would be difficult at best in a single cluster. One way
to move forward is by observing a larger sample of nearby clusters at
high Galactic latitude using RM-synthesis around 350~MHz. A fractional
bandwidth of 20\% is then sufficient to separate cluster emission from
Galactic foreground emission. If RM-synthesis does not reveal any
bubble-like polarized structures, similar to the ones towards the
Perseus-cluster \citep{DeBruynBrentjens2005}, it must be concluded
that the structures detected towards the Perseus cluster really belong
to the Galactic ISM. At this moment, however, there is not enough
evidence to draw a final conclusion.

\begin{acknowledgements} The Westerbork Synthesis Radio Telescope is
operated by \nobreak{ASTRON} (Netherlands Foundation for Research in
Astronomy) with support from the Netherlands Foundation for Scientific
Research (NWO).  \end{acknowledgements}


\bibliographystyle{aa}

\end{document}